%% file: draft.tex
\documentclass[referee]{jfm}
\usepackage{natbib}
\usepackage{amsmath}
\usepackage{graphicx}
\usepackage[usenames]{color}
\newcommand{\mb}[1]{\mathbf{#1}}

\newcommand{\bfb}[1]{\mbox{\boldmath $ #1 $}}
\newcommand{\beq}{\begin{equation}}
\newcommand{\eeq}{\end{equation}}
\newcommand{\rd}{\rm d}

\begin{document}
\title{Liquid interfaces in viscous straining flows: Numerical studies of the
selective withdrawal transition}

\author{
MARKO K. BERKENBUSCH$^1$\thanks{mkb@uchicago.edu},
ITAI COHEN$^2$\thanks{ic64@cornell.edu} \and
WENDY W. ZHANG$^1$\thanks{wzhang@uchicago.edu}
}

\affiliation{$^1$Department of Physics \& James Franck Institute, The
University of Chicago, 929 E. 57th Street, Chicago IL 60637,
USA\\[\affilskip]
$^2$Physics Department, Cornell University, Ithaca, NY
14853, USA}
\maketitle
\begin{abstract}
\input{abstract}

\end{abstract}
\section{Introduction}
\input{intro}
\section{Background}
\input{background}
\section{Modeling Selective Withdrawal}
\input{model}
\input{formulation}

\section{Results}
\input{overview_result}
\subsection{Interface Evolution \& Transition}
\input{result}
\subsection{Interface Evolution under Different Withdrawal Conditions}
\input{select}
\subsection{Comparison with Experiment}
\input{compare}

\section{Discussion}
\input{discuss}

\section{Conclusion}
\input{conclude}

\section*{Acknowledgements}
We thank Sidney R. Nagel and Sarah Case for encouragement and 
helpful discussions. We also acknowledge helpful conversations with
Francois Blanchette, Todd F. Dupont, Alfonso Ganan-Calvo,
Leo P. Kadanoff, Robert D. Schroll, Laura Schmidt, 
Howard A. Stone, Thomas P. Witelski,
Thomas A. Witten and Jason Wyman. This research was
supported by the National Science Foundation's Division of Materials
Research (DMR-0213745),
the University of Chicago Materials Lab (MRSEC), and by the DOE-supported ASC /
Alliance Center for Astrophysical Thermonuclear Flashes at the University of
Chicago.

\appendix
\section{}
\input{appendix}
\bibliography{stokes}
\bibliographystyle{jfm}
\end{document}

%% file: abstract.tex

This paper presents a numerical analysis of the transition from
selective withdrawal to viscous entrainment. In our model problem, an
interface between two immiscible layers of equal viscosity is deformed
by an axisymmetric withdrawal flow, which is driven by a
point sink located some distance above the interface in the upper layer. 
We find that steady-state hump solutions, corresponding to
selective withdrawal of liquid from the upper layer, cease to exist
above a threshold withdrawal flux, and that this transition
corresponds to a saddle-node bifurcation for the hump
solutions. Numerical results on the shape evolution of the steady-state 
interface are compared against previous experimental
measurements. We find good agreement where the data overlap. However,
the numerical results' larger dynamic range allows us to show 
that the large increase in the curvature of the hump tip near transition is not
consistent with an approach towards a power-law cusp shape, 
an interpretation previously suggested from inspection of the 
experimental measurements alone. Instead
the large increase in the curvature at the hump tip reflects a logarithmic
coupling between the overall height of the hump and the curvature at
the tip of the hump.

%% file: intro.tex

\begin{figure}
\centering
{
\label{fig:subfig:a} 
\includegraphics[width=1.5in]{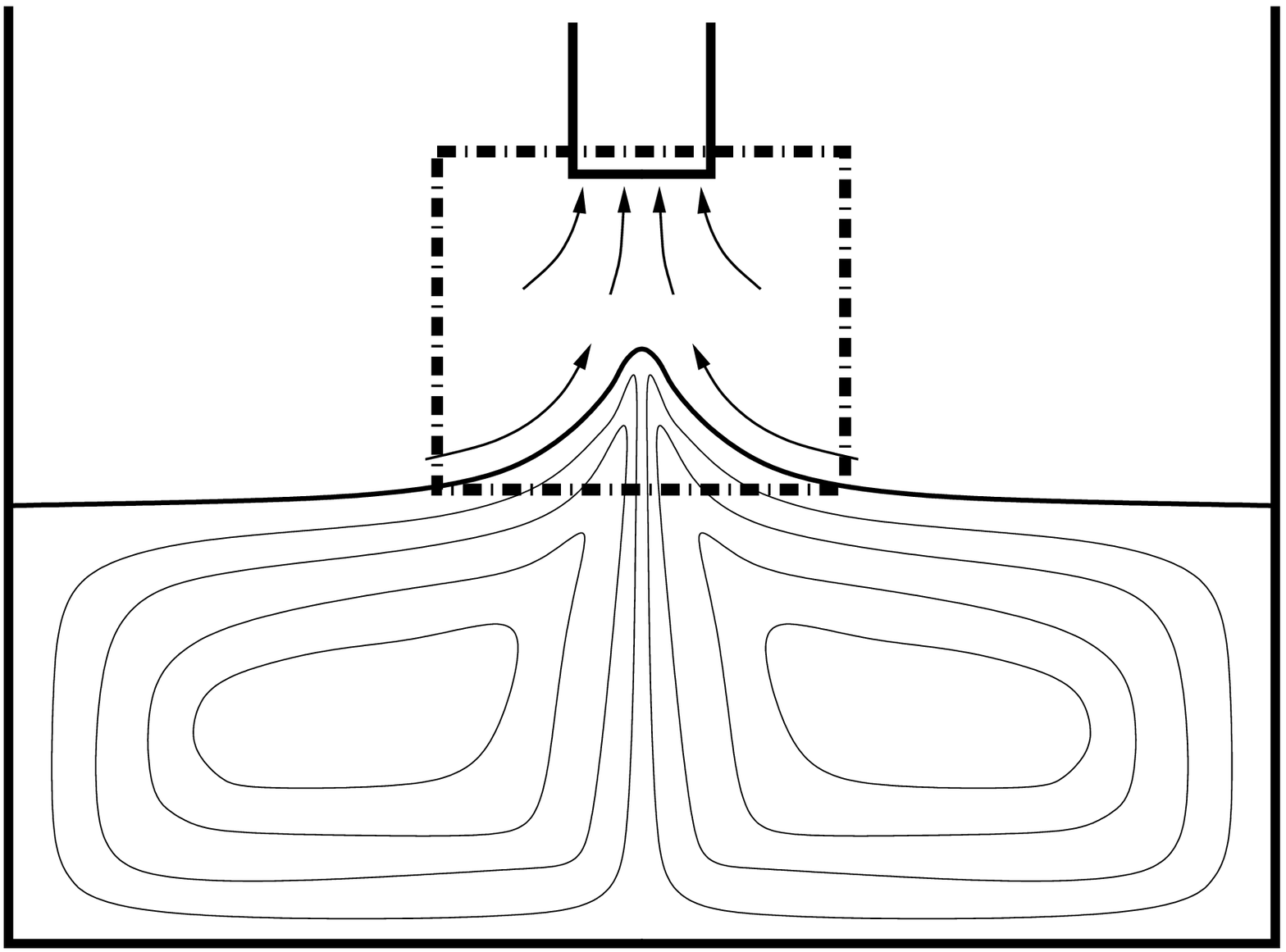}
}
\hspace{0.2cm}
{
\label{fig:subfig:b} 
\includegraphics[width=1.5in]{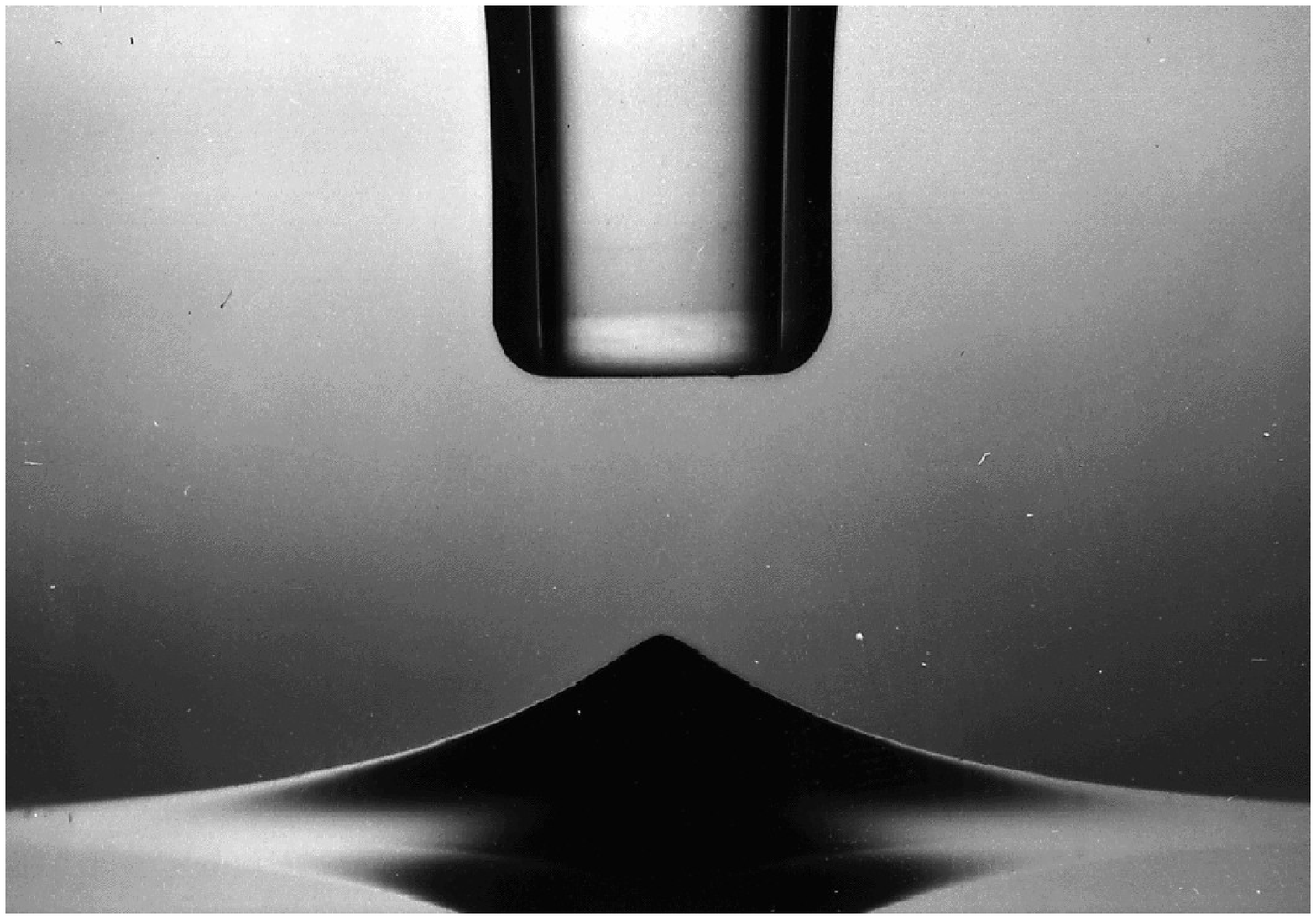}
}
\hspace{0.2cm}
{
\label{fig:subfig:c}
\includegraphics[width=1.5in]{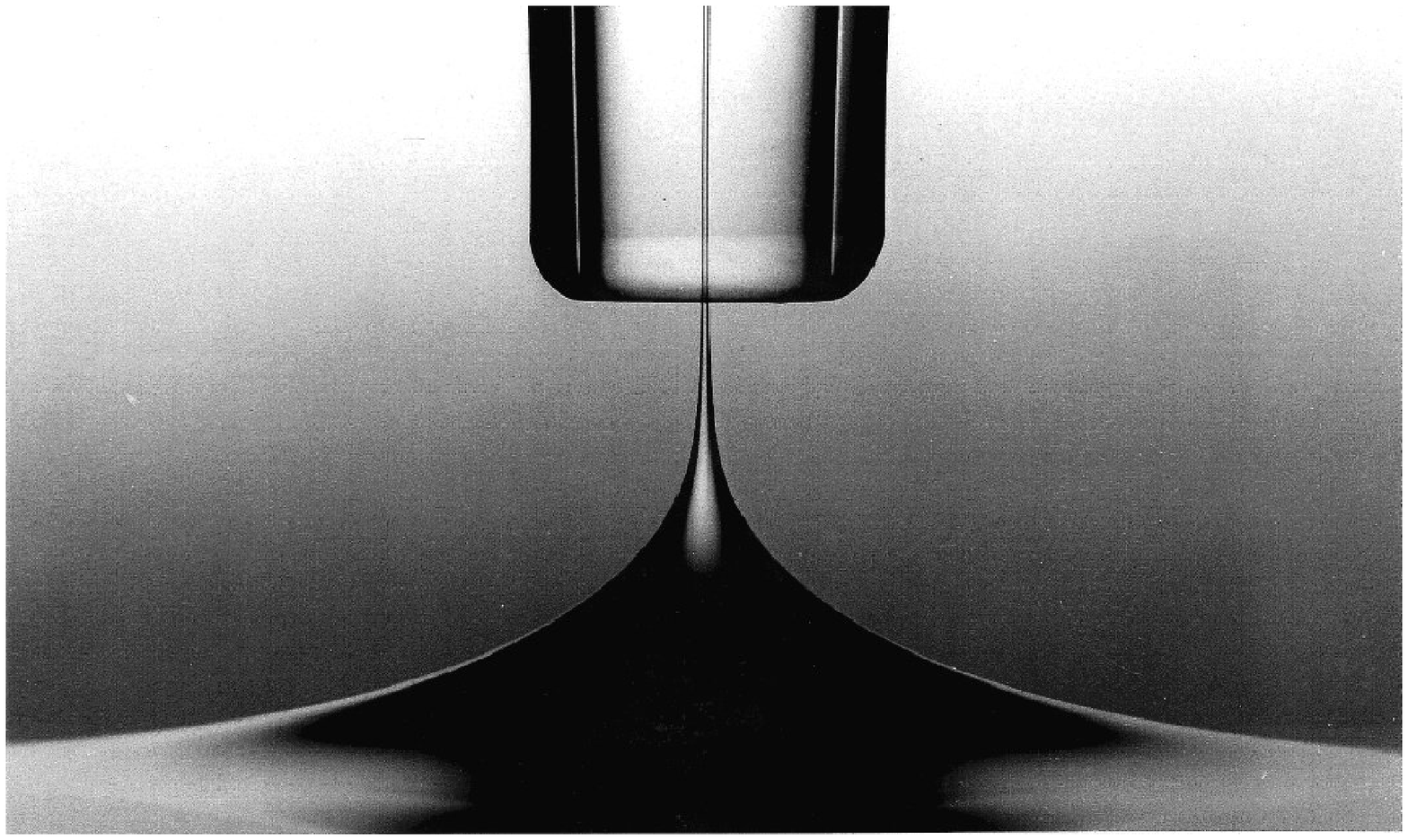}
}
\caption{
(a) Schematic of experiment. 
Liquid is withdrawn at a prescribed volume flux $\tilde{Q}$ through a
tube inserted into the top layer and replenished at the same rate. The top
layer is a viscous silicone oil. The bottom layer is a
mixture of water and glycerin. The two layers have comparable
viscosities. The flow rates are small so that viscous effects are
significant in both layers. The tube diameter is 1.6 mm. 
(b) Selective withdrawal regime: below $\tilde{Q}_c$ a
steady-state hump forms on the interface. (c) Viscous entrainment
regime: above $\tilde{Q}_c$ a
steady-state spout forms. Photos
courtesy of Cohen \& Nagel.
} 
\label{fig:exp} 
\noindent\rule{5.3in}{.5mm}
\end{figure}

Topological transitions of a fluid interface are a key step in many important 
and complex physical processes. Simple examples include the formation and coalescence 
of droplets where local stresses typically determine the fluid flows near the transition. 
Here, we examine a different scenario where the topological transition in the interface 
shape is driven by a large-scale steady-state flow.
Figure~\ref{fig:exp} depicts the experiment which 
inspired our numerical study. A deep layer of viscous silicone oil overlays a
second, immiscible layer, comprised of a mixture of water
and glycerin~(\cite{CohenNagel02}). Steady-state large-scale
flows are created in the two fluid layers by withdrawing liquid through a tube 
placed in the upper layer and replenishing the same amount of liquid
back into the upper layer at locations far away from the point of withdrawal. The rate 
of fluid withdrawal $\tilde{Q}$ dramatically affects the shape of the 
steady-state interface. When $\tilde{Q}$ is below a threshold value $\tilde{Q}_c$, 
the steady-state interface forms a hump. The lower-layer flow has the form of a toroidal
recirculation, with a stagnation point at the tip of the hump. 
Since only liquid in the upper
layer is withdrawn into the tube, 
this is known as the selective withdrawal regime.  
When the withdrawal flux $\tilde{Q}$ is larger
than $\tilde{Q}_c$, the flows and the interface remain steady, but 
the interface attains a different topology. The steady-state interface is
now ``open'' and has the shape of a spout which extends smoothly into
the tube. Since liquid is now withdrawn from both layers, this is
known as the viscous entrainment regime. In this regime, the amount of
liquid in the lower layer slowly decreases while the amount of liquid
in the upper layer slowly increases. The large-scale toroidal
recirculation in the lower-layer is weakly perturbed by the onset
of entrainment. The stagnation point is deflected slightly downwards,
from the hump tip to the interior. 

Analogous transitions from selective withdrawal to entrainment 
occur in many industrial and
physical processes. Some examples on large lengthscales include 
the intrusion of water into an oil reservoir during the last
stages of petroleum recovery~(\cite{renardy85,renardy85b}),
water-coning in sandy beds~(\cite{forbes04}), 
the formation of long-lived plumes by thermal convection in the earth's
mantle~(\cite{manga02}) and pollution recovery on lakes and
oceans~(\cite{imberger82}).  
On small lengthscales, the bursting of a liquid drop in a steady
straining flow~(\cite{Taylor34,Grace82,RallisonAcrivos78,Stone94}),
the deposition of a thin trailing tendril 
by a drop sliding down an incline~(\cite{limat04}) and the
formation of thin cylindrical tendrils and droplets in microfluidic
devices using axisymmetric flow-focusing~(\cite{anna03,Link04,Lorenceau05,utada05})
all involve topological transition of steay-state interfaces. 

In the selective withdrawal experiments, the curvature at the 
hump tip increases significantly
as the transition is approached from
below~\cite{CohenNagel02,Cohen04}. 
Intriguingly, measurements of
the hump height and curvature evolution near the transition suggest the 
steady-state hump shape evolves towards a power-law cusp. 
In order to explain this apparent divergence, Cohen and Nagel made an analogy 
to recent works on topological 
transitions that correspond to the formation of a finite-time
singularity in the governing equations. Some 
examples are the break-up of a liquid drop into two daughter
drops~(\cite{Eggers97,Cohen99, Zhang99,doshi03}), the coalescence of two
drops into one~(\cite{oguz90,egg99,thoroddsen05}), and the eruption of an  
electrohydrodynamic spout under the sudden application of a large
electric-field~(\cite{oddershede00}). In all these situations, the
topology change necessarily requires a divergence in the mean
curvature at a point on the interface. As a result, the Laplace
pressure due to surface tension, which is proportional to the mean
curvature, diverges as well, thereby creating a singularity in the
governing equations. This suggested analogy between the topological
transition in viscous withdrawal and finite-time singularity formation
is surprising. While the break-up of a liquid drop requires that the
interface deforms continuously from a single, connected shape to
several disconnected shapes, 
and therefore requires the formation of a singular
shape at break-up, the topological transition in this experiment deals
with changes in the {\it steady-state} interface shape. Therefore the
transition from selective withdrawal to viscous entrainment does not require
that the hump shape evolve into the spout shape continuously as a
function of the withdrawal flux $\tilde{Q}$. Generically, one would
expect a discontinuous change in the steady-state shape across the
selective withdrawal to viscous entrainment threshold. The apparently,
nearly continous shape-change observed in the experiment is,
therefore, rather unusual. 

This surprise motivates our work. Here we conduct
a numerical analysis of a model withdrawal
process in which the interface deformation is solely controlled by
viscous stresses and surface tension. We focus on the selective
withdrawal regime and analyze how the interface fails at the transition. 
In particular we are interested in elucidating the various factors that determine
the final hump shape. Experiments show that this final shape is well correlated 
with the minimum spout thickness attained in the viscous entrainment
regime (figure~\ref{fig:exp}c).  
Consequently, a detailed understanding of this topological transition
will help advance a variety of practical applications. One such application entails 
takeing advantage of viscous  
entrainment to encapsulate biological cells for transplant
therapy~(\cite{Cohen01a,Wyman04}).  
In this technique, the minimum spout size directly determines the
minimum thickness of a protective polymer coating that can be applied to the cells.

The rest of this paper is organized as follows: In section~2 we discuss
related works and background materials. In section~3 we describe the model 
withdrawal process and its numerical formulation for a system in which 
the upper and lower fluid viscosities are equal. In section~4 we analyse 
solutions to the model. Section~4.1 describes in detail the interface evolution near
the transition and the sharpening of the hump tip. We find that 
scalings of the hump curvature and hump height with the rate of withdrawal 
identify the transition as a saddle-node bifurcation. In accordance with 
this picture we find that the hump solution remains smooth as the 
transition is approached. Surprisingly, we also find that near the transition, 
the hump height couples logarithmically to the curvature at the hump tip. 
In section~4.2, we show that, for the equal viscosity layers investigated, 
these key findings are independent of the particular 
withdrawal conditions in the numerical simulations. In section~4.3, we compare 
the numerical and experimental results and find excellent agreement
between them.  
Moreover, we show that the unusual logarithmic coupling between the hump 
height and the hump curvature can easily lead to a misinterpretation that 
the system approaches a steady state singularity at the transition. 
A more detailed discussion of the distinction between the previous
interpretation and our results is presented 
in section~5. 

Our findings indicate the transition from selective withdrawal to 
viscous entrainment involves a discontinuous change in the
steady-state shape that has the structure of a saddle-node bifurcation. 
This discontinuity is effectively obscured in practice by the 
logarithmic coupling between the hump height and the hump curvature, 
which causes changes in the hump curvature to be far larger, and 
therefore more evident, than changes in the hump height as the 
transition is approached.

%% file: background.tex
Many different realizations of flow-driven topological transitions
have been investigated in previous works. 
Here, we review those studies that directly assess the ingredients 
necessary for a topological transition to take place and the
appropriate ways of characterizing the evolution of the interface near
such a transition. 

\subsection{Viscous entrainment in the absence of surface tension}

A number of experimental and theoretical works have focused on
the transition from selective withdrawal to viscous entrainment in
stratified, two-layer systems where surface tension effects are assumed
to be either negligible or weak. These studies were motivated by
geophysical flows, in particular the emptying of a magma chamber 
during a volcanic explosion. Blake \& Ivey investigated the threshold
withdrawal flux $\tilde{Q}_c$ necessary for the onset of entrainment
in miscible, stratified layers~(\cite{ivey85}). In a follow up numerical 
study~(\cite{Lister89}), Lister analyzed the onset of entrainment in two 
liquid layers of equal 
viscosity and found that a finite value for $\tilde{Q}_c$ exists only 
if surface tension effects, however small, are included in the analysis.  
These studies did not analyze the hump evolution near the transition.

\subsection{2-D analogues}
Air entrainment,
such as occcurs in high-speed coating~(\cite{simpkins00}) and in the
impact of a jet of viscous liquid into a layer of the same viscous
liquid~(\cite{Lorenceau04}) is often accompanied by the interface
approaching a two-dimensional power-law cusp
shape~(\cite{Joseph91}). Jeong \& Moffatt showed analytically that, in
the idealized limit where air flow effects are negligible, there is no
transition from selective withdrawal to air
entrainment~(\cite{JeongMoffatt91}). Instead the interface shape
approaches a steady-state singularity in the shape of a power-law cusp
as the flow rate is increased towards infinity.  More recent thereotical
and experimental analyses~(\cite{Eggers01,lorenceau03}) showed that,
when air flow effects are included, they perturb the idealized
dynamics so that a transition to air entrainment occurs at finite flow
rate. The approach towards a steady-state singularity is then cut-off at
a small lengthscale. Since the cut-off for air entrainment originates
from viscous stresses associated with air flow, the cut-off
lengthscale has a strong dependence on the viscosity contrast and
increases as $(\mu_0/\mu)^{4/3}$, where $\mu_0$ is the air
viscosity.  

\subsection{Selective withdrawal in 3D with viscous flow and surface tension}
In their studies of 3-D axisymetric selective withdrawal, Cohen \&
Nagel~(\cite{CohenNagel02}) 
used two liquids of comparable viscosities and measured
$\tilde{h}$, the hump height, and $\tilde{\kappa}$, the mean curvature at
the hump tip, as a function of the withdrawal flux $\tilde{Q}$. In a follow up 
study Cohen~(\cite{Cohen04}) analyzed the transition for different pairs of
liquids. In particular, the viscosity ratio, defined as the viscosity
of the liquid to be entrained divided by the viscosity of the
entraining liquid, was varied from $O(1)$ to $10^{-3}$. 
Analysis of the measurements show that, as the transition from
selective withdrawal to viscous entrainment is approached from below,
the steady state hump curvature increases dramatically. Cohen and Nagel 
showed that this rate of increase can be fit to a power law divergence. 
They also noted however, that the hump never reaches a cusp shape. Instead 
the final stage of the hump evolution is cut off by the transition. This 
behavior lead to the interpretation that a singular solution for the
steady-state shape organizes the 
nearly continuous transition between the steady-state hump and spout shapes. 
The transition from a hump to a spout was observed to occur when the radius 
of curvature at the hump tip was O($50$) $\mu$m or more. Moreover, the 
cut-off lengthscale shows little dependence on the viscosity contrast between the two 
layers. Consequently axisymmetric viscous withdrawal in 3D differes strikingly 
from air entrainment in 2D, which shows a strong dependence of the cut-off
lengthscale on the viscosity contrast. 

\subsection{Viscous drainage}
An analogous, but not necessarily equivalent, topology change occurs in viscous
drainage experiments, in which a viscous liquid exposed to air is
placed in a container with an opening at the bottom surface and is
allowed to drain out of the container due to its own
weight~(\cite{chaieb_preprint,sylvain06}).  As in the setup by Cohen
\& Nagel, the drainage creates a
large-scale toroidal recirculation whose viscous stresses deform the
interface. Thus the
drainage experiment resemble an upside-down version of 
the two-layer withdrawl experiment. A key difference 
is that the layer depth of the withdrawn fluid 
changes with time in the drainage experiment, 
but remains constant in the setup used by Cohen
\& Nagel.  As the liquid drains out and the layer depth decreases below a
critical value, a sharp cusp develops on the interface, a feature
identified by Correch du Pont \& Eggers as a steady-state singularity.  
A further decrease in the layer depth causes the cusp to
approach closer to the bottom surface and eventually intrude into the
aperture through which the fluid is draining. Correch du Pont \& Eggers 
report that there is no transition from selective withdrawal to viscous 
entrainment throughout the entire drainage process. Moreover,
the maximum curvature obtained in the drainage experiments, which
corresponds to using two fluid layers with a viscosity contrast of
$10^{-6}$, are far larger than the maximum curvature measured by Cohen
\& Nagel for two fluid layers with a viscosity contrast of
$10^{-3}$. The origin of these different outcomes is, at present, not understood. 

\subsection{Viscous withdrawal in 3D with surface tension}
Motivated by the thin stable spout created in the viscous
entrainment regime, Zhang analyzed the transition in the reverse direction, 
from viscous entrainment to selective withdrawal. The analysis focuses on the
regime where the viscosity of the fluid entrained is far smaller than
the viscosity of the entraining liquid, 
and constrains the spout shape to be nearly cylindrical
throughout~(\cite{Zhang04}). These 
simplifications allow the steady-state spout shape to be described
via a long-wavelength model. Surprisingly, results for the model show that, 
changing the boundary conditions on the interface, without changing
the material parameters, can profoundly change the nature of the shape
transition.  For some boundary conditions, the steady-state interface
deforms continuously with withdrawal flux 
from a spout to a hump across the transition. For other boundary
conditions, the steady-state interface changes discontinuously at the
transition. Thus, there appears to be a subtle interplay between the entraining 
flow and surface tension effects. As a result, the transition is 
strongly influenced by constraints on the large-scale shape of the
interface. Case \& Nagel have recently characterized the steady-state
spout shape experimentally~(\cite{case06}).

\subsection{Emulsification}
Another extensively studied shape transition that has strong points of 
similarity with the selective withdrawal to viscous
entrainment transition occurs when a single liquid drop is emulsified
into several drops by an axisymmetric straining 
flow~(\cite{Taylor34,RallisonAcrivos78,Grace82,Stone94}). In both situations, 
an imposed flow
far from the interface induces flows near the interface between two
viscous liquids. The steady-state interface is deformed at low
flow rates and ``broken'' at high flow rates. The crucial differences
are that, for drop emulsification, 
there is no steady-state interface above the threshold flow rate. The
drop is simply stretched out indefinitely by the flow. Also, even in
the analog of the selective withdrawal regime, when
a steady-state shape exists for the drop, the flow dynamics are
different because the flow inside the drop must assume a form
satisfying global volume conservation. In contrast, for the two-layer
withdrawal experiment, the layer depth is always much larger than the
hump height. It is then reasonable to expect that the global volume
constraint does not introduce a significant modification to the flow 
dynamics. We will show this expectation is indeed borne out by results 
from the numerics. 

The viscosity contrast between the drop
and surrounding fluid strongly affects the 
deformation-to-burst transition. When the drop
is far less viscous than the surrounding liquid, the steady-state
shape approaches a cusp shape as the burst transition is approached
from below. In particular, the radius of curvature at the two,
elongated ends of the liquid drop is cut-off
on an exponentially small
lengthscale~(\cite{AcrivosLo78}). Long-wavelength analyses of 
the extended drop shape in this regime also show that the burst
transition corresponds to a saddle-node bifurcation in the
leading-order solution for the overall
drop shape~(\cite{Taylor64,Buckmaster73}). When the drop
viscosity is comparable with the surrounding liquid, the steady-state
shape remains nearly spherical as the burst transition is
approached. Recently, numerics and phenomenological arguments have shown
that the shape transition when the drop viscosity is equal to the
surrounding liquid viscosity corresponds to a saddle-node 
bifurcation~(\cite{Navot99,BlawCrisLoew02}). 

\subsection{Inviscid selective withdrawal}
Finally, we note that a similar flow-driven topological transition
exists at high Reynolds number when 
water drains out of a filled bathtub~(\cite{sautreaux01,lubin67}). In practice,
this is always accompanied by the formation of a vortex. 
More generally, withdrawal from two stratified layers
of inviscid fluids is relevant for transport in water storage
reservoirs, where often a layer of fresh water overlies a layer of
salty water~(\cite{imberger82}). Intriguingly, 
some of the results for the idealized problem of withdrawal from two
layers of perfectly inviscid fluids~(\cite{tuck84,vanden87,miloh93})
suggest that, 
for the inviscid fluid system, the transition from selective
withdrawal to full entrainment corresponds to the formation of a
singular shape on the steady-state interface. This has motivated many recent
theoretical and numerical studies, despite the fundamental 
difficulty that, without the
small-scale smoothing provided by viscosity and/or surface tension in
real life, the idealized problem is fundamentally ill-posed and can
only be approached via careful limiting processes. For the same
reason, direct comparison with experiments has also been
difficult. For our study, 
the most relevant and suggestive results are that two-dimensional withdrawal 
differs significantly from axisymmetric
withdrawal~(\cite{forbes04,stokes05}). Also, in both 2D and
axisymmetric withdrawal, whether selective withdrawal or full
entrainment is realized can depend crucially on the initial
conditions which determine transient evolution of
the interface, instead of being solely determined by
the boundary conditions~(\cite{hocking01}).

In sum this review of viscous flow driven
topological transitions shows studies of flow-driven topological transitions
have produced a wealth of surprises. 
The wide range of behaviors observed indicate that apparently trivial differences 
in the exact experimental configurations can have a dramatic effect
on the transition structure. On the other hand, the interface
evolution observed in these different situations shares a number of
common features, such as the existence of a topological transition at
a finite flow-rate, and the development of small-scale features on the
interface near the transition. In order to make progress 
on the 3-D selective withdrawal problem, it is likely that experiments, 
simulations and theoretical calculations, all of which address the same 
flow geometry, will be needed. This is one reason why our study focuses on 
devising and analyzing a model withdrawal process that can be
directly compared with the previous experiments~(\cite{CohenNagel02}).

%% file: model.tex

This section describes our numerical model in the following order: 
we first identify the key parameters in the viscous
withdrawal experiments by Cohen \& Nagel and then describe an
idealized withdrawal problem which retains these key
features. A numerical formulation of the idealized problem, together
with the approximations, is given in section 3.1. Section 3.2 
gives details of the numerical implementation.  The key approximations
are that in our calculation the gradual flattening of
the liquid interface due to hydrostatic pressure on large lengthscales
is approximated by a hard-cutoff lengthscale $a$. 
The gradual decay of the induced flow in
the lower layer below the interface is approximated by a cut-off
condition requiring that the pressure in the lower layer becomes
uniform laterally across the liquid layer. 
Comparisons of results obtained using
these approximations against those calculated using more realistic
boundary conditions (section 4.2) and the experiment (section 4.3) 
show that the key features of the withdrawal are well reproduced by
the simple numerical model. 

In the experiment, the height of the hump created is at most a few
millimeters. The tube height $S_p$ is comparable in size, ranging from
$0.255 $ cm to $0.830$ cm. The capillary lengthscale $\ell_\gamma$, which 
characterizes the relative importance
of surface tension and hydrostatic pressure, is about $0.55$ cm. Deformations on 
lengthscales below $\ell_\gamma \equiv \sqrt{\gamma/\Delta \rho g}$ are
stabilized by surface tension. Deformations on lengthscales above
$\ell_\gamma$ are stabilized by hydrostatic pressure. All these
lengthscales are much smaller than the dimensions of the liquid layer 
which is $12$ cm deep and $30$ cm in extent. Also, 
when tube height is sufficiently large, the interface 
deformation changes very little when the tube diameter is
changed. This is the regime in which all the experimental data were
taken. 

Since the tube height $S_p$ and the capillary lengthscale
$\ell_\gamma$ are comparable, both lengthscales are expected to influence
the dynamics. As a result, there are several reasonable ways to define
the Reynolds number, and it is unclear how the Reynolds number of the
withdrawal flow should change with $S_p$ or $\ell_\gamma$. To avoid
this ambiguity, we use the observed hump 
height $\tilde{h}$ as the characteristic lengthscale and define the Reynolds
number $Re$ as $\rho \tilde{Q} \tilde{h} / (\mu 4 \pi S^2_p)$, where $\rho$ is the
density of the upper liquid, $\mu$ the viscosity of the upper
layer, and $\tilde{Q}/4 \pi S^2_p$ a typical flow rate. 
An estimate using parameters from the experiment 
yields $Re \approx 0.1$, indicating that inertial effects are small. 
The stress balance on the interface is therefore characterized by 
the capillary number, 
\beq
Ca = \frac{\mu}{\gamma}\frac{\tilde{Q}}{4 \pi S^2_p}
\eeq
a ratio of the strength of the viscous stresses exerted
by the flow relative to the Laplace pressure due to surface
tension. For a typical experiment, 
the transition from selective withdrawal to viscous
entrainment corresponds to a threshold capillary number $Ca_c$ in the range of
$10^{-3}$. Thus surface tension effects are strong even at the transition. 

\subsection{Minimal Model}

These results from the two-layer viscous withdrawal experiment
motivate the following idealization. Since the container size
is much larger than the characteristic lengthscale of the hump, the two
immiscible liquid layers are taken as infinite in extent and in
depth. We use a cylindrical coordinate system where
$z=0$ corresponds to the height of the undisturbed, flat interface and
$r=0$ is the centerline of the withdrawal flow. To drive the
withdrawal, we prescribe a sink flow
\beq
\mb{u}_\textrm{ext}(\mb{x}) = \frac{\tilde{Q}}{4 \pi | \mb{x}_S-\mb{x} |^2}
(\mb{x}_S-\mb{x})
\label{sink}
\eeq
where $\mb{x}_s = \tilde{S} \mb{e_z}$ corresponds to a sink placed at
height $\tilde{S}$ above the undisturbed interface and $\tilde{Q}$ is
the strength of the point sink. 
This choice allows $\tilde{S}$ to be the only lengthscale imposed on
the flow, consistent with the experimental observation that the hump 
shape depends primarily on the height of the withdrawal tube $S_p$ and not
on its diameter. 
Instead of explicitly 
including the effect of hydrostatic pressure, 
we require that the
interface is deformed by the withdrawal flow only within the region $0
\leq r \leq a$. The interface $S_I$ is pinned at $0$ deflection at
$r=a$. Beyond this point, the interface is taken to be entirely
flat. 
Physically this hard cut-off, or pinning length $a$ corresponds
roughly to the capillary lengthscale $\ell_\gamma$.
In the calculation the slope of of the interface at $r=a$ is adjusted
 so that the pinning condition is satisfied at all flow rates. This
introduces an error in the interface shape near $r=a$. 
This error is guaranteed to be small when 
the sink height $\tilde{S}$ is far smaller than $a$. It turns out to
be also small even when $\tilde{S}$ is comparable with $a$. (In the experiment,
tube height $S_p$ is either comparable with or smaller than the
capillary lengthscale $\ell_\gamma$.)

Finally, since the Reynolds number associated with flows in the
experiment is small and the interface evolution is observed to remain
essentially the same even as the layer viscosities are made unequal, 
we idealize the flow dynamics in both liquids to be purely viscous. We
also assume the two liquids have exactly the same viscosity. These
assumptions mean that the flow dynamics in the upper layer satisfies 
\begin{equation}
  \mb{\nabla} \cdot \bfb{\sigma}_1 = -\mb{\nabla}p_1 + \mu \mb{\nabla}^2
\mb{u}_1 = \mb{0}, 
  \quad \mb{\nabla}\cdot \mb{u}_1= \mb{0}
\end{equation}
where $\mb{u}_1$ is the disturbance velocity field created in the
upper layer due to the presence of the liquid interface, 
$p_1$ is the pressure and $\bfb{\sigma}_1$ is the stress field associated
with the disturbance velocity field $\mb{u}_1$. Note $\mb{u}_{\rm
ext}$ is a solution of Laplace's equation and therefore does not give
rise to a uniform pressure distrubtion. The
flow in the lower layer is created solely via the flow in the upper
layer past the interface. Therefore, $\mb{u}_2$, 
the disturbance velocity field in the lower layer, satisfies 
\begin{equation}
  \mb{\nabla} \cdot \bfb{\sigma}_2= -\mb{\nabla}p_2 + \mu \mb{\nabla}^2
\mb{u}_2= \mb{0}, 
  \quad \mb{\nabla}\cdot \mb{u}_2 = \mb{0}
\end{equation}
where $p_2$ is the pressure field in the lower layer, $\bfb{\sigma}_2$ the
fluid stress field in the lower layer. 

Since the flow dynamics in both layers are completely viscous, and
therefore described by the linear Stokes equations, we can 
represent the velocity field as an integral over a
suitably defined closed
surface, as discussed in earlier works and in textbooks on viscous
flow~(\cite{Lorentz07,Ladyzhenskaya63,pozrikidis92}). As a reminder,
the key result of the boundary integral formulation is that a Stokes
velocity field $\mb{u}$ at a point $\mb{x}$ is given by an
integral over any closed surface $S$ 
enclosing $\mb{x}$. The surface
integral has the form 
\beq
\mb{u}(\mb{x}) =  \int_{S} \bfb{J}(\mb{r}) \cdot 
  [\mb{n} \cdot \bfb{\sigma}(\mb{y})] \ \rd S_{\mb{y}} + 
  \int_{S} \mb{n} \cdot \bfb{K}(\mb{r}) \cdot \mb{u}(\mb{y}) \ \rd S_{\mb{y}} 
\label{basic_bint}
\eeq
where $\mb{y}$ is the point on the surface that you are integrating
over and $\mb{n}$ is an outward-pointing surface normal.
The tensors $\mb{J}$ and $\mb{K}$ are defined as:
\[
  \mb{J}(\mb{r}) =  \frac{1}{8\pi\mu} \left( \frac{\mb{1}}{r} +
\frac{\mb{rr}}{r^3} \right) \qquad
  \mb{K}(\mb{r})  =  -\frac{3}{4\pi} \frac{\mb{rrr}}{r^5}
\qquad \mb{r} = \mb{x}-\mb{y}.
\]
Physically, equation (\ref{basic_bint}) says that the velocity at the
interior point $\mb{x}$ can be expressed as a sum over two different
kinds of contributions over the the closed surface $S$. 
The first term on the right hand side of equation~(\ref{basic_bint})
corresponds to the component of fluid stress normal to the surface $S$ 
exerted by flows past the enclosing surface. The second contribution
corresponds to flows into and along the closed surface $S$ (second term
on the right hand side of equation (\ref{basic_bint})).  If 
the point $\mb{x}$ lies outside the volume enclosed, then the
contributions over the closed surface cancel, so that the right hand
side of (\ref{basic_bint}) sums to $0$. 
A point on the closed surface $S$ is a special case. When the closed
surface $S$ is continuous and smoothly varying, the velocity at
$\mb{x}$ is simply an average of the contribution from $S$ to $\mb{x}$
if it were in the exterior and the contribution from $S$ to
$\mb{x}$ if it were an interior point.  As a result, the velocity
for a point $\mb{x}$ on the surface $S$ can be written as 
\beq
\frac{1}{2}\mb{u}(\mb{x}) =  \int_{S} \bfb{J}(\mb{r}) \cdot 
  [ \mb{n} \cdot \bfb{\sigma}(\mb{y}) ] \ \rd S_{\mb{y}} + 
  \int_{S} \mb{n} \cdot \bfb{K}(\mb{r}) \cdot \mb{u}(\mb{y}) \ \rd S_{\mb{y}}. 
\label{bint_surface}
\eeq

\begin{figure}
  \centering
  \includegraphics[width=0.7\textwidth]{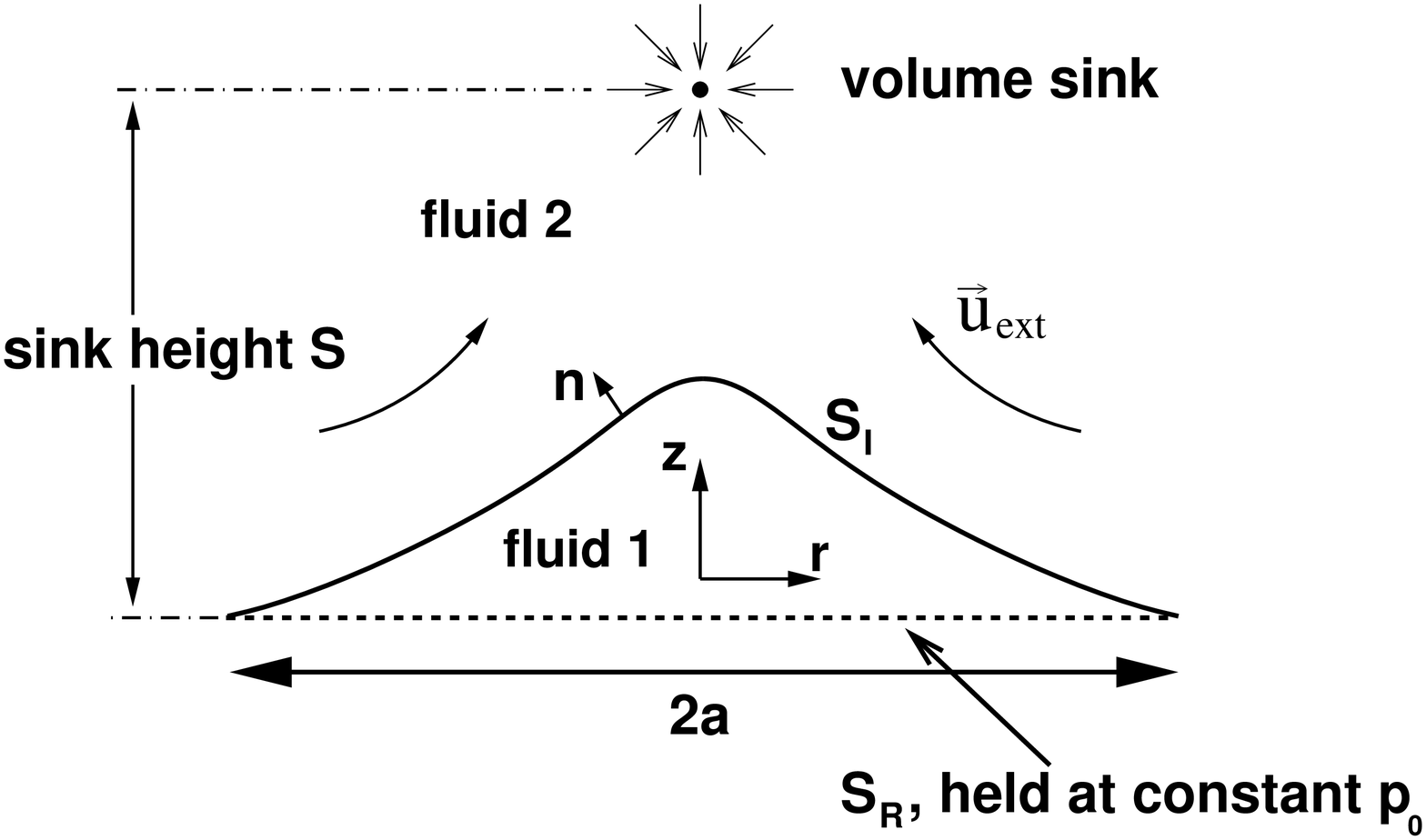}
  \caption{Simplified numerical model of selective withdrawal.  The
upper and lower liquid layer is separated by an interface $S_I$,
constrained so that the deflection is nonzero only within a radius
$a$. At a finite withdrawal flux, 
the surface $S_R$ lies entirely within the lower layer.}
  \label{fig:surface_185.49_clean}
\noindent\rule{5.3in}{.5mm}
\end{figure}

These results allow us to derive an integral equation for the
time-evolution of the interface in our idealized withdrawal problem.  To begin, we
note that we can enclose the entire upper layer as follows: first 
we define the surface $S_\infty$, which has the shape of
a hemispherical shell with radius $R_\infty$. As $R_\infty$ goes to
infinity, the surface $S_\infty$ and the liquid interface $S_I$
encloses all the liquid in the upper layer.  From (\ref{basic_bint}), 
we can write the disturbance velocity
$\mb{u}_1$ in the upper layer as 
\beq
 \mb{u}_1(\mb{x}) =  \int_{S_I + S_\infty} \bfb{J} \cdot 
  ( \mb{n} \cdot \bfb{\sigma}_1) \ \rd S_{\mb{y}}
  + \int_{S_I + S\infty} \mb{n} \cdot \bfb{K} \cdot \mb{u}_1 \ \rd
  S_{\mb{y}} . 
\eeq
Since $\mb{u}_1$ decays rapidly to $\mb{0}$ at infinity, the contribution
from the shell $S_\infty$ approaches $0$ as $R_\infty$
is taken to infinity. As a result, the only remaining contributions
are due to the liquid interface $S_I$, so that we can write the 
disturbance velocity $\mb{u}_1$ in the upper layer as  
\beq
\mb{u}_1(\mb{x}) =  \int_{S_I} \bfb{J} \cdot 
  ( \mb{n} \cdot \bfb{\sigma}_1 ) \ \rd S_{\mb{y}}
  + \int_{S_I} \mb{n} \cdot \bfb{K} \cdot \mb{u}_1 \ \rd
  S_{\mb{y}} 
\label{bint_upper_interior}
\eeq
From (\ref{bint_surface}), the velocity at a point on the interface is given by 
\beq
\frac{1}{2}\mb{u}(\mb{x}) =  \int_{S_I} \bfb{J} \cdot 
  ( \mb{n} \cdot \bfb{\sigma}_1 ) \ \rd S_{\mb{y}}
  + \int_{S_I} \mb{n} \cdot \bfb{K} \cdot \mb{u}_1 \ \rd
  S_{\mb{y}} 
\quad \textrm{ for } \mb{x} \in S_I.
\label{bint_upper_interface}
\eeq
Finally, consider a point $\mb{x}$ on the surface $S_R$ given by
$z=0$, as sketched in figure~~\ref{fig:surface_185.49_clean}. At a
finite withdrawal flux, the interface is deflected away 
from the $z=0$ plane. As a result, points on $S_R$ lie entirely inside
the lower layer and therefore are not enclosed by the surface $S_I +
S_\infty$, therefore the contribution from the surface integral over
$S_I + S_\infty$ vanish at a point $\mb{x}$ on $S_R$, or 
\beq
\mb{0} =  \int_{S_I} \bfb{J} \cdot 
  ( \mb{n} \cdot \bfb{\sigma}_1 ) \ \rd S_{\mb{y}}
  + \int_{S_I} \mb{n} \cdot \bfb{K} \cdot \mb{u}_1 \ \rd
  S_{\mb{y}} 
\quad \textrm{ for } \mb{x} \textrm{ in lower layer.} 
\label{bint_upper_exterior}
\eeq

Next we consider the volume of liquid in the lower layer 
enclosed by the closed surface comprised
of $S_R$ and $S_I$. Again, starting with (\ref{bint_surface}), we can
write the velocity at a point $\mb{x}$ on either the liquid interface
$S_I$ or the surface $S_R$, henceforth referred to as the ``reservoir
surface'', as 
\begin{eqnarray}
\label{comb_bint}
 \frac{1}{2} \mb{u}(\mb{x}) & = &  \int_{S_I} \bfb{J} \cdot 
  ( \mb{n} \cdot \bfb{\sigma}_2 ) \ \rd S_{\mb{y}}
  + \int_{S_I} \mb{n} \cdot \bfb{K} \cdot \mb{u}_2 \ \rd S_{\mb{y}} 
  + \int_{S_R} \bfb{J} \cdot 
  ( \mb{n} \cdot \bfb{\sigma}_2 ) \ \rd S_{\mb{y}} \\
  &+& \int_{S_R} \mb{n} \cdot \bfb{K} \cdot \mb{u}_2 \ \rd S_{\mb{y}}
\quad \textrm{ for } \mb{x} \in S_I \textrm{ or } S_R
  \nonumber 
\end{eqnarray}
where $\mb{n}$ again is an outward pointing surface normal. 
As a result, for a point $\mb{x}$ on the liquid interface $S_I$, 
the surface integral over $S_I$ in (\ref{comb_bint})
has exactly the opposite sign as the surface integral over $S_I$ in
(\ref{bint_upper_interface}). Since the velocity is continuous across
the interface, the two surface integrals involving the $\mb{K}$ tensor
cancel exactly when the two equations, (\ref{comb_bint}) and 
(\ref{bint_upper_interface}), for the velocity at
a point $\mb{x}$ on the interface are added together. As a result, the
velocity on the liquid interface can be re-written as 
\begin{eqnarray}
\label{bint_sI}
\mb{u}(\mb{x}) & = & \int_{S_I} \bfb{J} \cdot 
 [ \mb{n} \cdot \bfb{\sigma} ]_{-}^{+}\, \rd S_{\mb{y}} 
  + \int_{S_R} \bfb{J} \cdot ( \mb{n} \cdot \bfb{\sigma}_2 ) \ \rd
S_{\mb{y}} \\
  & + & \int_{S_R} \mb{n} \cdot \bfb{K} \cdot \mb{u}_2 \ \rd S_{\mb{y}}
\quad \textrm{ for } \mb{x} \in S_I \nonumber
\end{eqnarray}
where $[ \mb{n} \cdot \bfb{\sigma} ]_{-}^{+} = \mb{n} \cdot
\bfb{\sigma}_1 - \mb{n} \cdot \bfb{\sigma}_2$ 
denotes the jump in the normal stress across $S_I$ due to
surface tension and is equal to $2 \gamma \kappa \mb{n}$ where
$\kappa$ is the mean surface curvature and $\mb{n}$ points outwards
from $S_I$ and $S_R$. Similarly, adding
(\ref{bint_upper_exterior}) and (\ref{comb_bint}) yields an expression
for the velocity at a point $\mb{x}$ on the reservoir surface $S_R$
\begin{eqnarray}
\label{bint_sR}
  \frac{1}{2} \mb{u}_2(\mb{x}) & = & \int_{S_I} \bfb{J} \cdot 
 [ \mb{n} \cdot \bfb{\sigma} ]_{-}^{+}\, \rd S_{\mb{y}} \\
  & & + \int_{S_R} \bfb{J} \cdot (\mb{n} \cdot \bfb{\sigma}_2)\ \rd S_{\mb{y}}
  + \int_{S_R} \mb{n} \cdot \bfb{K} \cdot \mb{u}_2 \ \rd S_{\mb{y}}
\quad \textrm{ for } \mb{x} \in S_R. \nonumber
\end{eqnarray}
Equations (\ref{bint_sI}) and (\ref{bint_sR}) together provide an
expression 
for the velocity on the liquid interface as a function of the
interface shape, and $\mb{n} \cdot \bfb{\sigma}|_{S_R}$, the 
normal stress exerted by flow in the lower layer on the
reservoir surface $S_R$. In a full calculation,
$\mb{n} \cdot \bfb{\sigma}|_{S_R}$ would need to be obtained
separately and then used in (\ref{bint_sI}) and (\ref{bint_sR}) to
calculate the velocity on the liquid interface. Here, we employ a
drastic simplification and simply prescribe the normal stress
distribution on $S_R$. Specifically we require that the normal stress
is a spatially uniform pressure of size $p_0$, so that 
\beq
\mb{n} \cdot \bfb{\sigma}_2|_{S_R} = p_0 \mb{n} \ .
\label{p0}
\eeq
This choice of the stress condition (\ref{p0}) is motivated by the
observation that, since the lower layer is taken to be infinitely
deep, the stress distribution far from the interface must decay
smoothly onto one that corresponds to a stagnant layer. This is simply a
spatially uniform pressure field. By imposing this
distribution at $S_R$, instead of a boundary condition far from the
interface, we are essentially approximating the smooth decay that
would be obtained in a deep lower layer as an abrupt cut-off at
$S_R$. This is in the same spirit as approximating the effect of
hydrostatic pressure by introducing a pinning length $a$ for the
interface and turns out be just as effective in capturing the key
features of the interface evolution. A more realistic simulation
which includes a lower layer of finite depth, instead of imposing
(\ref{p0}) at $S_R$ yields essentially the same results (section 4.2). 
 
We are now ready to specify an evolution equation for the liquid
interface. The disturbance velocity on the interface is given by 
\begin{eqnarray}
\label{usI}
  \mb{u}(\mb{x}) & = & \int_{S_I} \bfb{J} \cdot 
  \mb{n} \left[ \gamma 2 \tilde{\kappa} \right]\, \rd S_{\mb{y}}\\
  & & + \int_{S_R} \bfb{J} \cdot \mb{n} \ p_0 \, \rd S_{\mb{y}}
  + \int_{S_R} \mb{n} \cdot \bfb{K} \cdot \mb{u}_2 \,\rd S_{\mb{y}} 
\quad \textrm{ for } \mb{x} \in S_I \nonumber
\nonumber
\end{eqnarray}
while the velocity on $S_R$ needed to evaluated the third surface
integral in (\ref{usI}) is given by (\ref{bint_sR}). The total
velocity on the interface is a sum of the disturbance velocity and the
imposed withdrawal flow. Therefore, to update the interface, we use
the kinematic condition
\begin{equation}
  \label{eq:kinematic_bc}
  \frac{d\mb{x}}{dt} = \mb{u}(\mb{x}) + \mb{u}_\textrm{ext}
\quad \textrm{ for } \mb{x} \in S_I \
  . 
\end{equation}
To find a steady-state hump profile for a given withdrawal flux, 
we start with an initial guess and solve (\ref{bint_sR}), 
(\ref{usI}), and (\ref{eq:kinematic_bc})
in succession  
until either a steady-state is obtained or the interface deformation
grows so large that the interface reaches the sink. 

Finally we nondimensionalize the withdrawal problem
via the following characteristic length-, velocity- and stress-scales
\beq
\hat{\ell} = a \qquad \hat{u} = \frac{\gamma}{\mu} 
\qquad \hat{p} = \frac{\gamma}{a} 
\label{nondim}
\eeq
The dimensionless withdrawal flux is then 
\beq
  Q = \frac{\mu \tilde{Q}/a^2}{\gamma}
\eeq
and is equivalent to a capillary number. 

%% file: formulation.tex
\subsection{Numerical Implementation}

We use a C++ code to solve the governing
integral equations~(\ref{bint_sR}), (\ref{usI}), and (\ref{eq:kinematic_bc}). 
Since the problem is axisymmetric,
all the surfaces involved can be represented as effectively
one-dimensional objects by their $r$ and $z$ coordinates in a cylindrical
coordinate system. 
The liquid interface is parameterized by mesh points $r_j\left(s_j\right),
z_j\left(s_j\right)$, where $s_j$ denotes the arclength along the surface
measured from the tip. The interface between mesh points is
approximated by a cubic spline. This way, the curvature calculated for the
Laplace pressure term varies continuously between mesh points. The
distribution and number of points is adapted to the geometric shape of
the surface. The algorithm we implemented chooses the density of
points to be proportional to the mean curvature (in most runs $\Delta
s = 0.05/\kappa$, where $\Delta s$ is the 
arclength between two adjacent mesh points). In addition we require
that adjacent mesh points cannot be separated by $\Delta s$ larger
than a maximum value. For most runs, this value was chosen to be of the
order of 1/20 in  dimensionless units. Since the interface at $Q=0$ is
nearly flat, it is represented by 20 grid points. During a typical run,
the number of mesh points on the interface is dynamically increased to about 60
in order to resolve a dimensionless tip curvature of approximately
75. Doubling the mesh resolutions yields no significant change in the
results.

The constant pressure reservoir surface, $S_R$, 
in the setup are represented by a uniform grid of points. The
velocity between points are interpolated linearly. The constant
pressure reservoir surface in the setup is represented by a mesh of 30
uniformly 
spaced points. For numerical reasons, it is simpler to work with a
small but non-zero 
pressure. For most runs shown in this paper, $p_0$ was chosen to be
$0.01$. As described in section 4.2, 
changing $p_0$ results in little relative changes in the results. 

The azimuthal component of the boundary integral equations
is integrated analytically and the resulting elliptic integrals are evaluated
numerically (see \cite{LeeLeal82}). 
The boundary integrals are evaluated using Gaussian quadratures.
Singular parts of the integral are dealt with separately using finer
subdivisions of the integration interval, and the resulting algebraic equations
for the unknown velocities and stresses are solved by LU (lower and upper
triangular matrix) decomposition.

The mesh points on the interface are advanced in time via 
explicit forward Euler time stepping. To ensure that the mesh points
remain evenly distributed, only the normal component of the velocity
is used to update the interface.
The length of each timestep is chosen to be
proportional to the smallest mesh spacing. The validity of this approach was
tested by doubling the time resolution for a given run and comparing to the
original results. Generally, results obtained this way differed by less than
1\%. We also experimented with an implicit backward Euler scheme,
but the evaluation of the necessary Jacobian was prohibitively slow, and the
results obtained by the two methods did not differ significantly. 

The scheme for finding steady state shapes and critical flow rates
follows the experimental procedure: starting with a static
configuration at zero flow rate, $Q$ is increased in small steps and
for each step the interface is allowed to relax into a steady state
shape. Steady state is detected by 
observing that the modulus of the highest normal velocity on the interface,
$\mb{u}_\textrm{max}\cdot \mb{n}$, falls below a certain threshold  ($10^{-3}$
in dimensionless velocity units in most cases; lower thresholds were tested
without significantly changing the results). When $u_\textrm{max}$ increases
monotonically by a certain factor, we deduce that a steady-state hump shape has
ceased to exist. In this case, the interval between the last steady state flow
rate and the current flow rate is divided in half, creating nesting intervals
containing the critical flow rate $Q_c$. This procedure stops 
when the size of the interval between $Q$ values has decreased below
the desired accuracy. To resolve the steady-state evolution near
$Q_c$, we also choose $Q$ values between those chosen from the
bisection and solve for the steady-state interface shape at these
values. 

Since the successive increases in the imposed withdrawal
flux $Q$ becomes very small near $Q_c$ and the initial velocity on the
interface is proportional to the increment in $Q$, the velocity on the
interface becomes small near $Q_c$ regardless of whether the interface
shape is stable. We therefore allow the interface to evolve for a longer
interval in time before checking whether the $u_\textrm{max}$ exceeds
the threshold value when $Q$ is close to $Q_c$. We have also compared
our results against runs done with an even longer interval of waiting
time to ensure that the simulation has converged onto the steady-state
solution.

%% file: overview_result.tex
Using the equations governing the model withdrawal problem described 
in section~3, we calculate numerical solutions for the interface evolution 
near the transition. Section~4.1 describes how the
curvature $\kappa$ and the hump height $h$ scale with the withdrawal flux 
$Q$ near the transition. This section also addresses the coupling between 
$h$ and $\kappa$. Section~4.2 shows these results are robust and insensitive 
to variations in the withdrawal conditions. Finally, in section~4.3, the 
numerical results are compared with those in the experiments. 

%% file: result.tex

\begin{figure}
  \centering
  \includegraphics[width=0.65\textwidth]{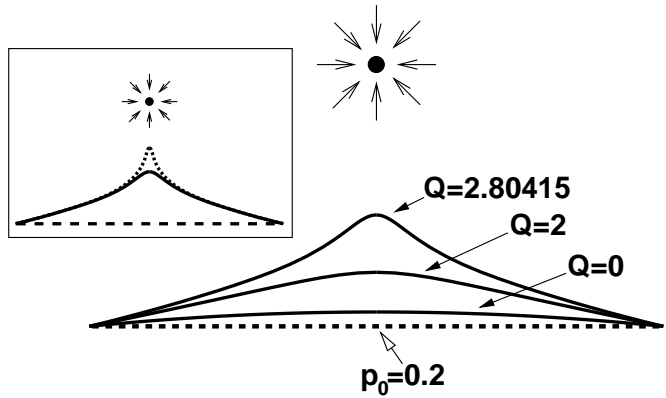}
  \caption{
Steady-state interface shapes (solid lines) at different $Q$. For
  $Q=0$, the interface assumes a spherical cap shape for which the
  Laplace pressure balances the reservoir pressure $p_0$. As $Q$
  increases, the steady-state shape develops a hump towards the sink.  
  The inset shows that, above the transition withdrawal flux $Q_c$,
  the unsteady interface (dashed line) develops a finger reaching 
  towards the sink. Here the dimensionless sink height $S \equiv \tilde{S}/a =
  0.2$ and the dimensionless reservoir pressure jump $p_0$ is $0.2$.}
  \label{fig:surface}
\noindent\rule{5.3in}{.5mm}
\end{figure}

Time-dependent simulations of equations(\ref{bint_sR}), 
(\ref{usI}), and (\ref{eq:kinematic_bc}) show that  
steady-state solutions corresponding to hump shapes 
are found only below a threshold flow rate $Q_c$. Above $Q_c$, no
hump solutions are found. Figure~\ref{fig:surface} shows a sequence of
the calculated steady-state shapes. At $Q=0$, the static interface is
a spherical cap shape determined by a balance of surface tension and
reservoir pressure $p_0=0.2$. When the imposed withdrawal flux
$Q$ is small, the interface is weakly perturbed from the spherical cap
shape. At larger $Q$, a broad hump
develops on the interface. As $Q$ approaches $Q_c$ the hump height
increases slightly but the hump curvature increases
dramatically. Above $Q_c$ the interface develops a finger-like
structure which lengthens over time and eventually reaches the
sink (figure 3 inset). For all $Q$, the shape of the hump at its tip
remains smooth,  
well-fit by a spherical cap with mean curvature $\kappa$.
We identify the disapparance of a steady-state hump solution at
$Q_c$ in the numerical solutions with the hump-to-spout transition
observed in the experiment.  

\begin{figure}
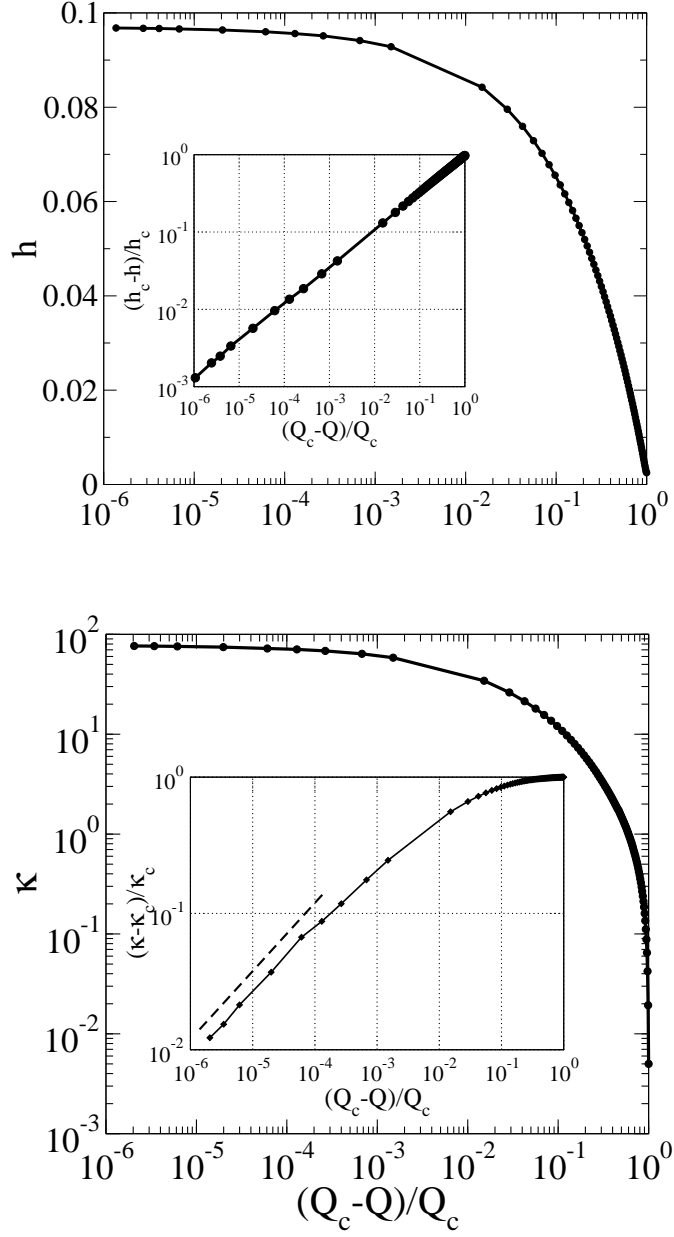

\centering
{
\includegraphics[width=0.65\textwidth]{figures/height_dq.eps}
}
\vspace*{0.2in}
\centering
{
\label{fig:kdq} 
\includegraphics[width=0.65\textwidth]{figures/curv_dq.eps}
}
\caption{Evolution of the calculated 
steady-state hump shape with withdrawal flux
$Q$. (a) Hump height $h$ vs. $Q_c-Q/Q_c$. The inset shows
$(h_c-h)/h_c$  approaches saturation as $\sqrt{(Q_c-Q)/Q_c}$. (b) Mean
curvature $\kappa$ at hump tip versus $Q$. The inset 
shows $(\kappa_c - \kappa)/\kappa_c$ approaches saturation as 
$\sqrt{(Q_c - Q)/Q_c}$. The dashed line depicts a square root powerlaw. 
} 
\label{fig:hkdq} 
\noindent\rule{5.3in}{.5mm}
\end{figure}

In the rest of this subsection, we will analyze one typical set of
results, obtained with dimensionless sink height $S=0.2$ and reservoir
pressure $p_0$ set to $0.01$. 
Hump solutions obtained for different dimensionless sink height $S
\equiv \tilde{S}/a$, $p_0$ values, 
and under different realizations of two-layer selective withdrawal 
can be found in section 4.2. We can 
quantitatively characterize the interface evolution by 
plotting the hump height $h$ and the mean curvature
$\kappa$ at the tip of the hump as functions of the withdrawal flux
$Q$ (figure~\ref{fig:hkdq}). As 
$Q$ approaches $Q_c$, the hump height saturates at $h_c$
(figure~4a) and the curvature saturates at $\kappa_c$. 
(figure~4b). The insets show that $h$ and $\kappa$ approach
their saturation values as 
\beq
\frac{h_c - h}{h_c} \propto \sqrt{\delta q} \qquad 
\frac{\kappa_c - \kappa}{\kappa_c} 
\propto \sqrt{\delta q} \qquad 
\delta q \equiv \frac{Q_c - Q}{Q_c}. 
\label{square_root}
\eeq
To obtain the best power-law fit, 
the $h_c$ and $Q_c$ values were adjusted slightly. 
The best fit is obtained when the values of $h_c$ and $Q_c$ 
were increased from the $h$ and $Q$ values corresponding to 
the final steady-state hump shape calculated by 
$0.03 \%$ and $4 \cdot 10^{-5}\%$ respectively. 
The same $Q_c$ value is used in the 
$\kappa_c - \kappa $ plot and the $h_c -h$ plot. 
The $\kappa_c$ value is adjusted by less
than $0.001 \%$ from the final hump curvature value obtained
numerically. All the 
adjustments are within the error bars of the simulation.

The observed square-root dependence of $h_c - h$ and $\kappa_c - \kappa$ 
suggests that the transition from
selective withdrawal to viscous entrainment occurs via a collision at
$Q_c$ of two steady-state hump solutions, one stable and one
unstable. In other words, the transition from selective withdrawal to
viscous withdrawal has the structure of a saddle-node bifurcation. Typically,
saddle-node bifurcations are associated with solutions that remain smooth 
at the bifurcation point. Therefore the observed square-root scaling
suggests that the final evolution of the interface as the transition is
approached is not organized by an approach towards a singular
steady-state shape. 

\begin{figure}
\centering
{
\label{fig:twoshapes} 
\includegraphics[width=0.4\textwidth]{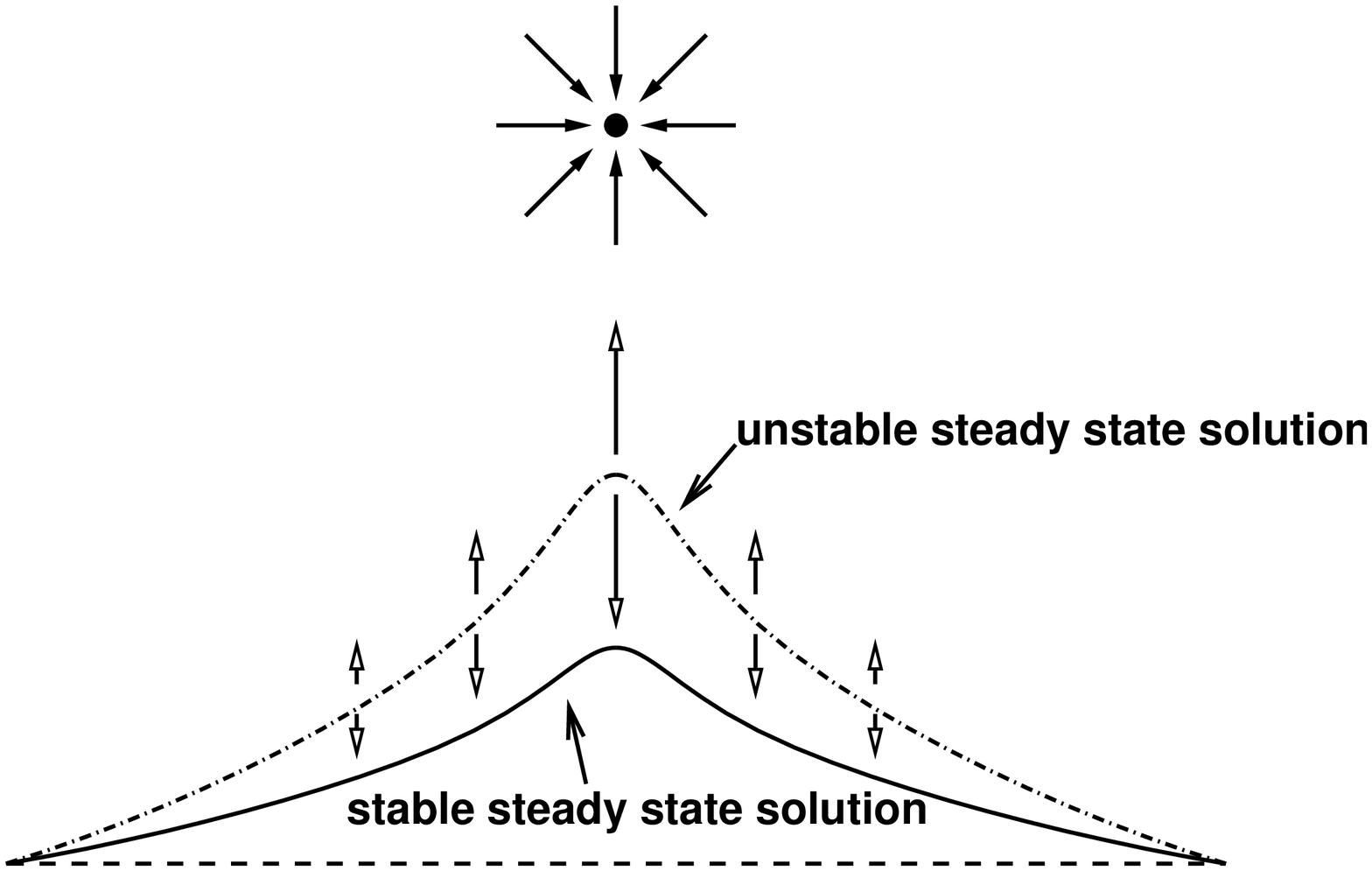}
}
\hspace{0.2cm}
{
\label{fig:merge} 
\includegraphics[width=0.4\textwidth]{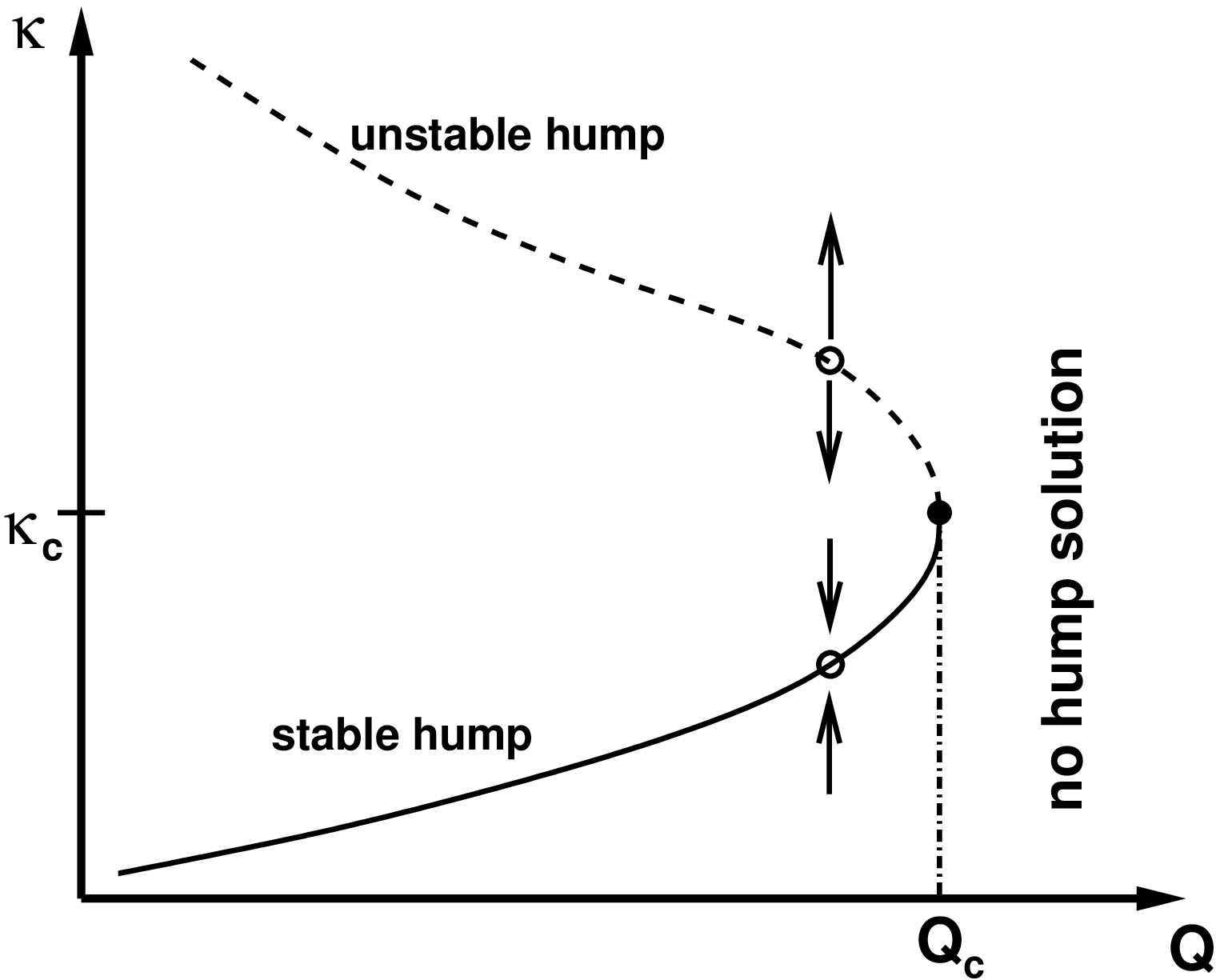}
}
\caption{
(a) Stable and unstable hump solutions in steady-state
selective withdrawal. (b) 
Saddle-node bifurcation diagram illustrating
the evolution of the hump curvature in the selective withdrawal regime. 
} 
\label{fig:saddle} 
\noindent\rule{5.3in}{.5mm}
\end{figure}

The saddle-node structure of the transition can be rationalized by considering 
the flow in the upper layer converges along the centerline and
therefore speeds up 
away from the tip of the hump. Since the hump solution is realized in a
time-dependent simulation and also observed in the experiment, it is
linearly stable. We therefore expect that small upwards perturbations of the
hump tip simply decay downwards as the hump shape relaxes towards the 
steady-state solution over time. However, if the upwards perturbation 
is very large, so that the
perturbed hump tip now lies very close to the sink, then surface tension
effects will be too weak to pull the perturbed interface
downwards. Instead, the hump tip will be drawn into the sink. This
suggests there exists an intermediate critical shape perturbation that neither
decays nor grows upwards but simply remains steady over time, as illustrated in
figure~\ref{fig:saddle}a. This critical shape perturbation would correspond
to an unstable hump solution. 

Figure 5b illustrates the evolution of the hump curvature as a
function of the flow rate for stable and unstable hump solutions. 
At low withdrawal flux values, only a very
large shape perturbation can cause the interface to be drawn into the
sink, we therefore expect the stable and the unstable hump shapes 
to be widely separated. This means the unstable hump must lie close to
the sink and, as a result, has a very curved tip. At moderate
withdrawal flux values,  
the perturbation size required to destabilize the hump solution
becomes smaller, implying that the unstable and stable solutions now
lie closer to each other. Concurrently, the curvature of the stable solution
increases, and the curvature of the unstable solution decreases. 
Near transition, even a small perturbation causes the interface to become
unstable and grow towards the sink. This suggests that the two
solutions lie very close each other and are nearly identical. 
Therefore it is very reasonable to expect the two solutions to become
identical, or coincide, at $Q_c$, thereby bringing about a saddle-node
bifurcation of the hump solution. The square-root scaling corresponds
to a smooth merging of the two solutions, so that the $\kappa(Q)$ 
curve at $Q_c$ has a shape of a parabolic lying on its side. 

\begin{figure}
\centering
{
\includegraphics[width=0.7\textwidth]{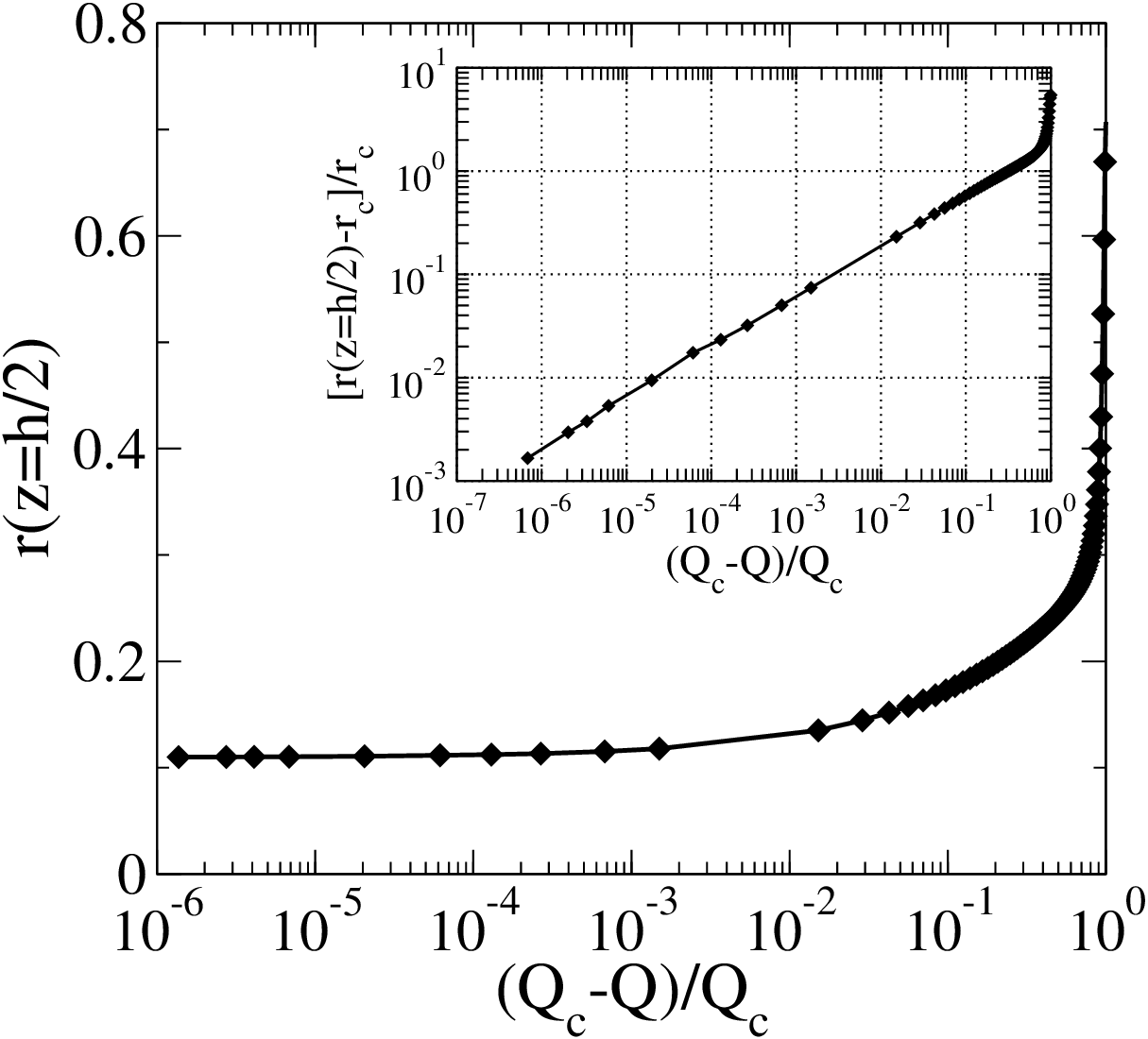}
}
\caption{Evolution of the hump radius at $z=h/2$ as a function of the
withdrawal flux $Q$. The inset shows how the radius at half height
reaches its saturation value.}
\label{fig:halfwidth}
\noindent\rule{5.3in}{.5mm}
\end{figure}

The scaling dynamics associated with how the hump solution saturates
as $Q$ approaches $Q_c$ is consistent with a saddle-node
bifurcation at $Q_c$. This is a generic mechanism for transition for
one type of solution from another in a nonlinear problem. 
However, the quantitative analysis in 
figure~\ref{fig:hkdq} shows one surprising and atypical feature:  the 
hump height and curvature saturate towards the final scaling behavior at very 
different $\delta q$ values. Specifically, the square-root scaling in 
$h_c-h$ is evident throughout its evolution, even at large $\delta q$. 
In contrast, the square-root scaling for $\kappa_c - \kappa$ becomes 
evident only when $\delta q$ has decreased below $10^{-3}$. This behavior 
suggests that the hump shape does not approach the transition 
uniformly. To test this idea we plot the hump radius at $z=h/2$ as a function 
of $Q$ in figure~\ref{fig:halfwidth}. We find that the radius at the half-height 
saturates to the final scaling behavior later than $h$, but earlier than $\kappa$ 
as $\delta q$ approaches $0$. 
This behavior indicates that, as $Q$ approaches $Q_c$, the overall shape of the 
hump, e.g. the hump height or its lateral extent, saturates first, followed by 
features on smaller lengthscales. The shape of the hump at its tip, which
corresponds to a feature on the smallest lengthscale, saturates last. 
This cascade of events is more typical of an approach towards a singular
shape, in which the evolution of features on different lengthscales
are nearly decoupled, so that features on smaller lengthscales saturate later than
features on large lengthscales. 

\begin{figure}
  \centering
  \includegraphics[width=0.7\textwidth]{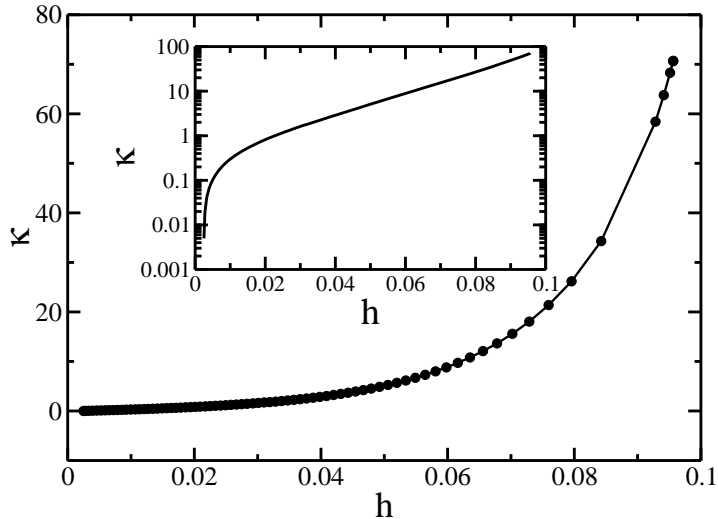}
   \caption{Mean curvature at the hump tip versus the hump height. Inset shows
the same data on a semi-log plot. When the deflection is sufficently large,
the hump height $h$ has a linear dependence on $\ln(\kappa)$.}
\label{fig:tctl_unrescale}
\noindent\rule{5.3in}{.5mm}
\end{figure}

To get some insight into this unusal hybrid character of the interface
evolution near transition, we plot $\kappa$ as a function of $h$ in 
figure~\ref{fig:tctl_unrescale}. At $Q=0$, 
the interface shape is given by a balance of Laplace pressure and
reservoir pressure $p_0$. For 
$p_0 = 0.01$, the shape is a nearly flat spherical cap. 
When the interface is only weakly perturbed from the $Q=0$ shape, both
$\kappa$ and $h$ increase linearly with $Q$, and therefore 
$\kappa \propto h$. 
This linear regime persists until $h \approx 0.02$. Beyond
this point, $\kappa$ increases much more rapidly
than $h$. Remarkably, this large increase of $\kappa$ relative to the
increase in $h$ is well-approximated by a simple mathematical expression,
as evident from the inset for figure~\ref{fig:tctl_unrescale}. In
the nonlinear deflection regime, 
\beq
\ln ( \kappa ) \propto h \ . 
\label{exp_form}
\eeq
This logarithmic coupling indicates that in the nonlinear regime, 
the hump solution evolves so that small increases in hump height 
corresponds to large increases in the hump curvature. This does not
correspond to a nearly continuous transition, as suggested by Cohen \&
Nagel (\cite{CohenNagel02}), since the hump curvature $\kappa$ never
decouples from the hump height. Instead, our results show that
$\kappa$ remains weakly coupled to $h$ and does not diverge as the 
tranisiton is approached. To gain some insights into the origin of this unusual 
coupling, we next investigate how changes in the withdrawal
conditions, in particular the sink height $S$ and the reservoir
pressure $p_0$, affect this relation between the hump curvature and
the hump height.  

%% file: select.tex

Naively one may expect that a logarithmic coupling between the hump
height and the hump curvature would be 
easily perturbed by changes in the boundary conditions, since
logarithmic coupling is often observed at the transition between 
different kinds of power-law scaling as some system parameter is
varied. In addition the withdrawal process analyzed numerically
is highly idealized and may therefore give results which are not
generic.  To address these concerns, we analyze 
numerical results for the steady-state interface shapes under
a variety of withdrawal conditions. 
Surprisingly, we find the logarithmic coupling~(\ref{exp_form}) to be 
a robust feature of equal-viscosity selective withdrawal in all the
numerical solutions. 

\begin{figure}
  \centering
    \includegraphics[width=0.7\textwidth]{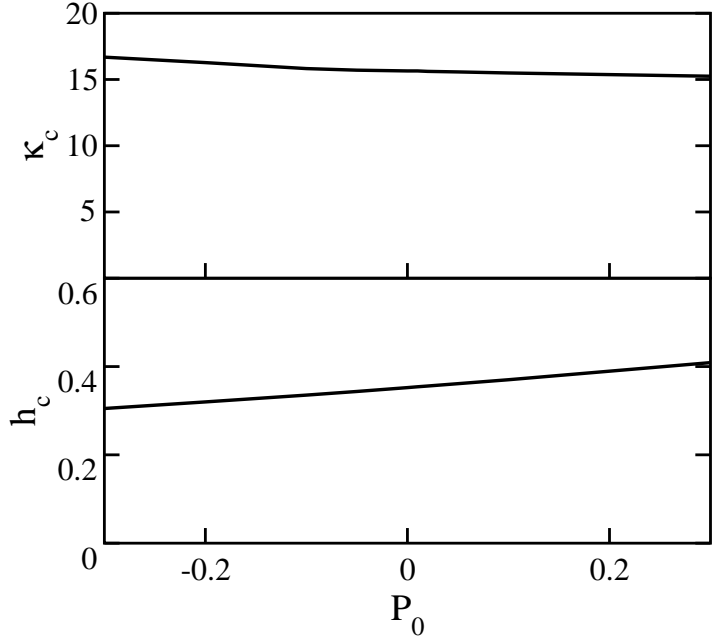}
      \caption{Calculated hump curvature $\kappa_c$ and hump height
$h_c$  at transition as a function of the reservoir pressure
$p_0$. The sink height $S=1$.}
\label{fig:diffp0}
\noindent\rule{5.3in}{.5mm}
\end{figure}

To determine the effect of the reservoir pressure $p_0$ on the 
transition structure we analyzed the numerical solutions for values 
of $p_0$ ranging from $-0.3$ to $0.3$. Larger $p_0$ values were 
not investigated because they result in static
shapes whose height is greater than the sink height even at $Q=0$. 
In figure~\ref{fig:diffp0} we plot $\kappa_c$ and $h_c$, the hump curvature and
hump height at transition versus $p_0$. We find that the
precise value of $p_0$ makes little difference to 
the values of $\kappa_c$ and $h_c$. Also, all the runs show the same logarithmic 
dependence of $\kappa$ on $h$ near the transition. 

\begin{figure}
\centering
{
\includegraphics[width=0.75\textwidth]{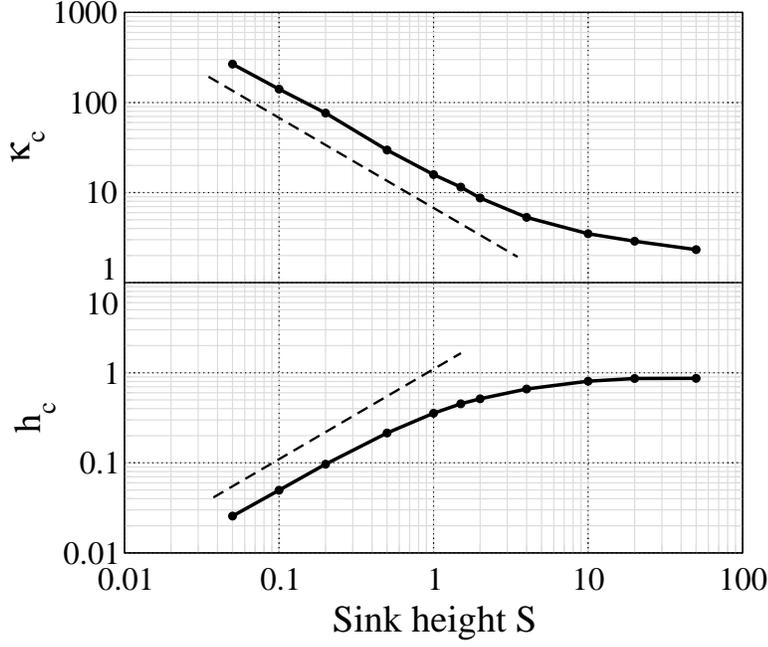}
}
\caption{Calculated hump curvature $\kappa_c$ and hump height
$h_c$  at transition different sink heights $S$. The reservoir
  pressure $p_0=0.01$. The
dashed line in the $\kappa_c$ vs. $S$ plot shows a $1/S$ dependence. The dashed
  line in the $h_c$ vs. $S$ plot shows a $S$ dependence.}
\label{fig:max_v_s} 
\noindent\rule{5.3in}{.5mm}
\end{figure}

In contrast, figure~\ref{fig:max_v_s} shows that varying the sink height $S$ strongly affects 
$\kappa_c$ and $h_c$. As the dimensionless sink height 
$S$ becomes much smaller than $1$, or $\tilde{S} \ll a$, 
the hump curvature at transition, $\kappa_c$, increases as $1/S$
while $h_c$ decreases as $S$. In the opposite limit, as $S$ becomes much
larger than $1$, or $\tilde{S} \gg a$, both $\kappa_c$ and $h_c$
approach constant values. These trends can be understood by 
analyzing the limiting cases where 
the sink is either very far or very near to the interface.
When the sink is placed very close to the interface, 
the effect of the pinning at $a$
becomes irrelevant since the interface is already flat at distances 
$O(S)$ where the flows become very weak. Consequently, $h_c$ and $\kappa$ only depend on $S$ 
when $S \ll 1$. In the opposite limit, where $\tilde{S} \gg
a$, the flow  at the interface becomes 
almost uniform because the lengthscale over which the imposed
withdrawal flow varies is much larger than $a$. Under these conditions 
the interface shape is primarily determined by the pinning condition at $a$ and
is insensitive to changes in the sink height. 

Figure~\ref{fig:max_v_s} also shows that the variation in
$\kappa_c$ tracks the variation in $h_c$ almost perfectly as $S$ is varied. 
Consequently, even though driving the withdrawal with a sink closer to the
interface produces a larger hump curvature at transition, it does not 
prduce a sharper hump tip. This is because the hump height,
and similarly the overall deformation at transition, is proportionally 
smaller as well, so that the product $\kappa_c \cdot h_c$
characterizing
the relative separation of lengthscale
between the hump tip and the hump height remains roughly constant
with $S$. 
Changing the sink height
primarily changes the absolute lengthscale of the
steady-state deformation at the onset of entrainment. It does not
change the shape of the deformation significantly.  

\begin{figure}
\centering
{
\includegraphics[width=0.75\textwidth]{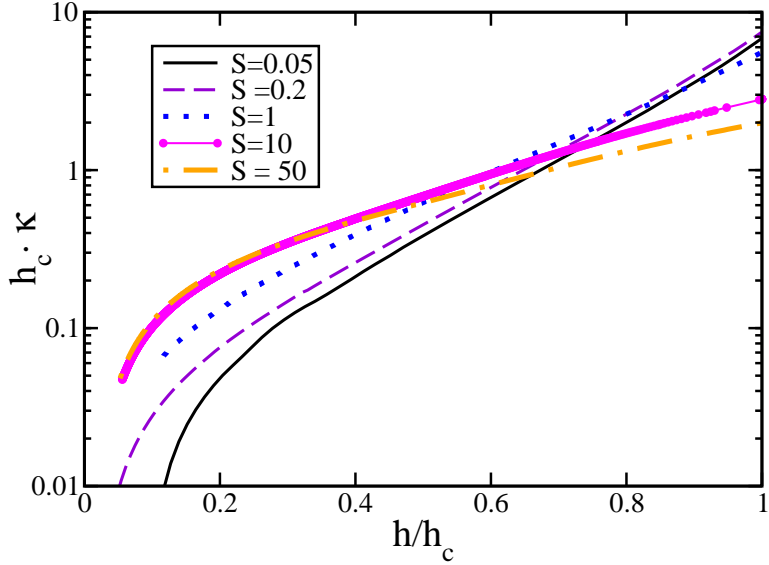}
}
\caption{Calculated 
hump curvature $\kappa$ versus hump height $h$ at $S=0.05$,
$0.2$, $1$, $10$ and $50$. 
} 
\label{fig:diffS} 
\noindent\rule{5.3in}{.5mm}
\end{figure}

This observation suggests that dividing all lengths by $h_c$ 
should scale out most of the variation observed at different sink 
heights. In figure~\ref{fig:diffS} we plot $h_c \cdot \kappa$ versus $h/h_c$ 
obtained from numerical solutions where $S=0.05$, $0.2$, $1$, $10$ and $50$ representing a $10^3$
variation in the sink height. We find that the curves are indeed brought close together by
this rescaling. For example, while the actual value of $\kappa_c$
changes by a factor of $100$ between $S=0.05$ and $S=50$, the rescaled
value $h_c \cdot \kappa_c$ changes by less than a factor of $4$. We do however, 
observe small changes in the slope and intercept of the logarithmic curves 
even when $\kappa$ and $h$ are rescaled. Nevertheless, from these results, we
conclude that the logarithmic coupling is a robust feature of our
idealized model of two-layer withdrawal. Changing 
either the boundary condition parameter $p_0$, or the forcing 
parameter $S$ within the simple model of withdrawl produces no qualitative
change in the evolution of the steady-state hump shape as a function
of the withdrawal flux.  

\begin{figure}
  \centering
{
  \includegraphics[width=0.7\textwidth]{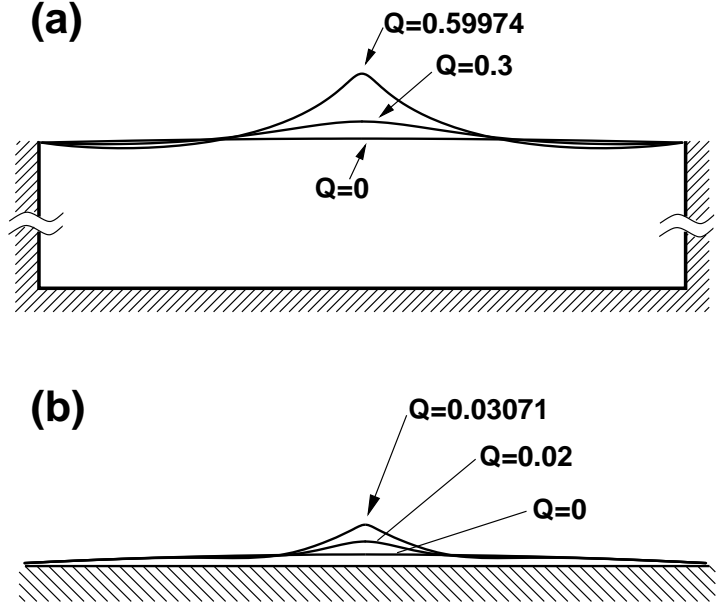}
}
  \caption{(a) Steady-state interface shapes obtained in selective
  withdrawal from a lower layer of finite depth. The cell depth equals
  the cell radius $a_1$ and the sink height $\tilde{S}/a_1 = 0.5$. (b)
  Steady-state interface shapes obtained in selective withdrawal from
  a thin lower-layer overlying a solid plate of radius $a_2$. The sink
  height $\tilde{S}/a_2 = 0.2$.}
  \label{fig:surf_seq_other}
\noindent\rule{5.3in}{.5mm}
\end{figure}

Given this robustness with respect to variation in system parameters
in the simple model, we next 
analyze the steady-state interface obtained under different
realizations of selective withdrawal. We find that even these 
different realizations, which correspond to different boundary
conditions and/or imposed withdrawal flows, 
fail to produce a qualitative change in
how the steady-state hump shape evolves. 
Figure~\ref{fig:surf_seq_other} shows the
interface shapes obtained in two different sets of calculations, each
with a lower layer of finite depth, instead of the infinite depth
assumed in the model calculation. 
In the first case, the lower layer
is contained in a cylindrical cell of radius $a_1$ and depth
$a_1$ (figure~\ref{fig:surf_seq_other}a). This corresponds to a
situation where the lower-layer is confined equally in both the radial
and vertical direction. The toroidal recirculation
established in the lower-layer is therefore much smaller in extent,
thereby resulting in an $O(1)$ change in the viscous stresses exerted
on the interface by the flow in the lower layer.

For this realization, 
the boundary integral formulation uses a closed surface comprised of
the liquid interface $S_I$, the side walls of the container
$S_\textrm{side}$ and the bottom wall of the containter $S_b$. The
disturbance velocity $\mb{u}$ on the interface is given by 
\begin{eqnarray}
\label{eqn:container_interface}
  \frac{1}{2} \mb{u}(\mb{x}) & = & \int_{S_I} \bfb{J} \cdot 
  \mb{n} \left[ \gamma 2 \tilde{\kappa} \right]\, \rd S_{\mb{y}}
  + \int_{S_\textrm{side}} \bfb{J} 
\cdot \mb{n} \ \bfb{\sigma}_\textrm{side}\, \rd S_{\mb{y}} \\
  & + & \int_{S_\textrm{b}} \bfb{J} 
\cdot \mb{n} \ \bfb{\sigma}_\textrm{b}\, \rd S_{\mb{y}} 
\quad \textrm{ for } \mb{x} \in S_I \nonumber
\end{eqnarray}
where $\mb{y}$ is the point on the surface that you are integrating
over, $\bfb{\sigma}_\textrm{side}$ and $\bfb{\sigma}_\textrm{b}$ correspond to the
normal stress exerted by the flow on the side wall and bottom wall of
the container. The two stresses are found by solving the equation
\begin{eqnarray}
\label{eqn:container_walls}
    \mb{0} & = & \int_{S_I} \mb{J} \cdot 
  \mb{n} \left[ \gamma 2 \tilde{\kappa} \right]\, \rd S_{\mb{y}} 
   + \int_{S_\textrm{side}} \mb{J} \cdot \mb{n} \cdot
\bfb{\sigma}_\textrm{side} \, \rd S_{\mb{y}} \\
  & + & \int_{S_\textrm{b}} \bfb{J} 
\cdot \mb{n} \cdot \bfb{\sigma}_\textrm{b}\, \rd S_{\mb{y}} 
\quad \textrm{ for } \mb{x} \in S_\textrm{side} \textrm{ or }
S_\textrm{b}. \nonumber
\end{eqnarray}
This corresponds to enforcing no-slip boundary conditions on the side
walls and the bottom walls of the container.

\begin{figure}
 \centering
{
\includegraphics[width=0.7\textwidth]{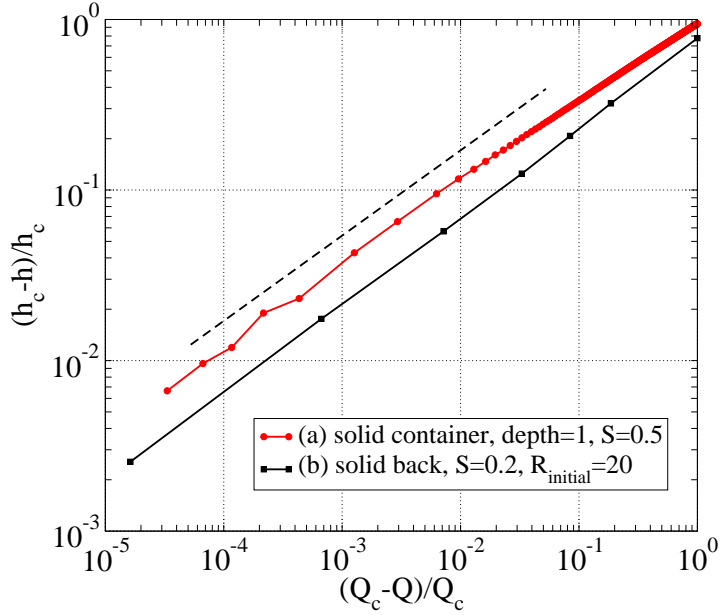}
}
\caption{Scaling of the 
calculated hump height $h$ near the transition flow rate for
two-layer withdrawal. Set (a) corresponds to
withdrawal from a cylindrical container. Set (b) corresponds to withdrawal from
a thin lower layer which is resting on a solid plate.}
\label{fig:tl_v_q_container_solid}
\noindent\rule{5.3in}{.5mm}
\end{figure}

The interface shapes shown are the calculated results 
and the cell radius $a_1$ is used as a characteristic lengthscale in 
nondimensionalizing various quantities. As the withdrawal flux $Q$
increases and the hump height increases, liquid is drawn from the
lower layer into the hump. Since volume in the lower layer is 
conserved, gathering extra liquid into the hump forces a shallow dip 
in the interface shape to develope some distance away from the
centerline. Thus, 
the hump profile no longer flattens monotonically
with $r$, in contrast to results obtained assuming an infinitely deep
lower layer. However, these qualitative changes to the overall shape
do not significantly change how the interface evolves with $Q$. The shape 
transition still corresponds to a saddle-node bifurcation
(figure~\ref{fig:tl_v_q_container_solid}). A
comparison of the rescaled $h_c \cdot \kappa$ versus $h/h_c$ curve
obtained for a cell of 
finite depth and the analogous curve obtained assuming an infinitely
deep layer in figure~\ref{fig:var} show the same trend. 

Figure~\ref{fig:surf_seq_other}b shows what happens when we change the
condition in the lower layer more drastically. In this set of
calculation, a thin 
layer of the lower layer liquid overlies a solid plate. The plate
radius $a_2$ is used as a characteristic lengthscale in the
nondimensionalization. The liquid is
assumed to wet the solid substrate so that the plate is covered 
by liquid from the lower-layer 
at all $Q$. The formation of a hump causes most
of the liquid in the layer to drain into the hump, leaving only a very
thin layer covering the rest of the solid plate. In the selective
withdrawal regime, the withdrawal flow requires liquid in the lower
layer to recirculate, moving along the interface towards the hump and
then along the lower solid surface away from the hump. When the lower
layer is thin, the viscous resistance associated with the steady-state
recirculation is amplified dramatically. In this
realization, therefore, the stress-balance on the steady-state
interface is strongly perturbed from that obtained for withdrawal with
an infinitely deep lower layer. 

For two-layer withdrawal with a thin layer of the lower liquid, 
the closed surface used in the boundary integral formulation consists of 
the liquid interface $S_I$ and the bottom wall of the containter
$S_b$. The equation for the 
velocity on the interface has the same form as~(\ref{eqn:container_interface})
except that the terms associated with the side wall surface
$S_\textrm{side}$ are absent. The normal stress on the solid surface
$\sigma_\textrm{b}$ satisfies equation (\ref{eqn:container_walls}),
without the sidewall contribution. 

Comparing against results from the simple model and 
results for withdrawal from a finite container, we can see that
reducing the thickness of the lower layer has an effect on the
steady-state shape of the interface obtained. The hump
shape is significantly more conical, and the flattening of
the interface at large $r$ results primarily from volume conservation,
rather than pinning at $a_2$ or the decay of the imposed withdrawal
flow away from the sink. However, as evident from
figure~{\ref{fig:tl_v_q_container_solid}} and figure~{\ref{fig:var},
the qualitative features are 
unchanged. The transition still corresponds to a saddle-node
bifurcation. The evolution of the hump height $h$ retains a
logarithmic coupling to the hump curvature $\kappa$. The main 
difference between withdrawal from an infinitely deep layer and
withdrawal from a very thin layer overlying a solid plate is a shift
in the slope of the rescaled $\kappa$ versus $h$ curve. 
\begin{figure}
\centering
{
\includegraphics[width=0.75\textwidth]{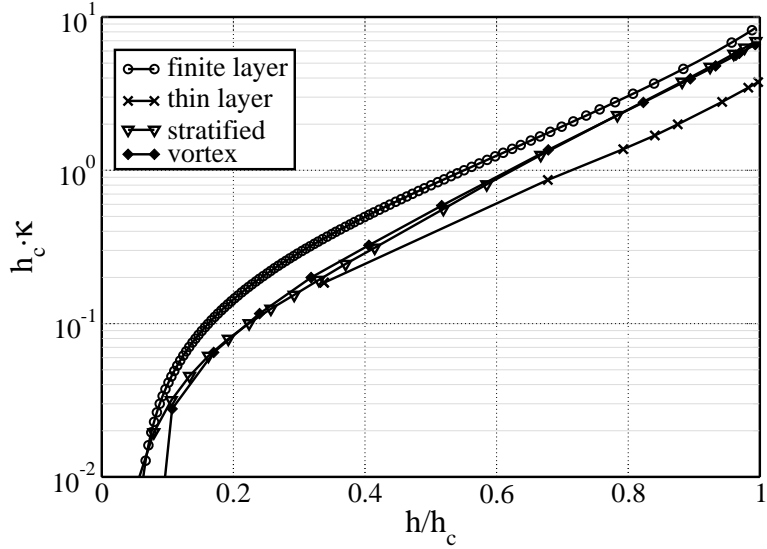}
}
\caption{Curvature versus hump height under different
withdrawal conditions. 
Open circles correspond to withdrawal with a
lower layer of finite depth contained in a cylindrical cell. The cell
depth equals the cell radius $a_1$ and the sink height $\tilde{S}/a_1
= 0.5$. The crosses correspond to withdrawal of a thin layer of 
liquid overlying a solid plate of radius $a_2$ and sink height
$\tilde{S}/a_2 = 0.2$. The open triangles correspond to withdrawal
from a stratified layer. The capillary lengthscale $\ell_\gamma/a =
0.1$ and the rescaled sink height is $0.2$. The solid diamonds correspond to 
withdrawal driven by a vortex ring of radius $a_\omega = a$, 
dimensionless strength $\omega$ and height $\tilde{S}/a = 0.2$. 
The $\kappa$-$h$ curve obtained for $S=0.2$ and $p_0=0.01$ (solid
line) is also shown for comparison. 
} 
\label{fig:var} 
\noindent\rule{5.3in}{.5mm}
\end{figure}

We have also conducted a series of
calculations which include the effect of hydrostatic pressure 
explicitly. This requires that we change the form of the normal stress
jump $[ \mb{n} \cdot \bfb{\sigma} ]_{-}^{+}$ across the liquid
interface $S_I$ in the J-integral of 
equations (\ref{bint_sR}) and (\ref{usI}) to 
\beq
[ \gamma 2 \tilde{\kappa} - \Delta \rho \mb{g} \cdot \mb{y}] \mb{n}
\eeq
where $\Delta \rho$ is the density difference between the two layers,
$\mb{g}$ the gravitational acceleration and $\mb{y}$ the location of a
point on the liquid interface.
Choosing the capillary lengthscale $\ell_\gamma$ to be
less than the pinning radius $a$ allows the interface to flatten out
at large radial distances due to stratification, as occurs in the
experiment, instead of due to the decay of the withdrawal flow or due
to surface tenson effect. Results for $\ell_\gamma/a = 0.1$ and $0.2$
show that introducing stratification explicitly
results in slight qualitative changes in the hump shape but the same trend in 
the rescaled $\kappa$ versus $h$ curve (figure~\ref{fig:var}). Finally, 
to assess the influence of details of the geometry 
of the withdrawal flow on the logarithmic coupling (\ref{exp_form}), 
we changed the imposed withdrawal flow from a sink flow
(\ref{sink}), to that associated with a vortex ring of size $a_\omega
= a$ and strength $\omega_0$. The axisymmetric velocity field 
is given in terms of a stream function $\phi$ 
\beq
u_r = \frac{1}{r} \frac{\partial \phi}{\partial r} \qquad
u_z = -\frac{1}{r} \frac{\partial \phi}{\partial z} 
\eeq
and the stream function has the form 
\beq
\phi (r, z) = \frac{1}{4 \pi} \int \int r r^\prime 
\omega(r^\prime, z^\prime) 
\rd z^\prime \rd r^\prime \int^{2 \pi}_0 \frac{\cos \theta \ \rd \theta} 
{\sqrt{(z - z^\prime)^2 + r^2 + r^{\prime 2} - 2 r r^\prime \cos \theta}}
\eeq
where the vorticity distribution $\omega(r^\prime, z^\prime)$ has the
form $\omega_0 \delta(a_\omega-r)\delta(S-z^\prime)$ for a vortex ring.  
This is a better approximation of the withdrawal flow
imposed in the experiment, both because there is no net volume
withdrawn and because the finite size of the vortex ring mimics the
finite diameter of the tube. The rescaled $\kappa$ versus $h$ curve obtained
from numerical solutions for withdrawal driven by a vortex ring again
has the same logarithmic coupling between the hump height and the hump
curvature near the transition.

In all the different realizations of selective withdrawal, the
evolution of the steady-state interface approaches the form
(\ref{exp_form}) for large deformations. 
As a result, $h_c \cdot \kappa_c$, the 
relative separation of lengthscales characterizing the hump shape at
transition, changes little even when the forcing and/or the boundary
conditions are changed drastically. In other words, regardless of what
type of flow we use to drive the transition, the degree of 
viscous resistance from flow within the lower layer, or whether the
interface flattens out on large lengthscale due to stratification or
surface tension effects, the hump shape at transition never becomes
significantly sharper than that first calculated in the simple
model. We conclude from this that the interface shape at $Q_c$
produced in viscous withdrawal with two liquid layers of equal viscosity
cannot be brought closer to a singular shape via changes in the
boundary conditions. This
contrasts sharply with theoretical results on viscous
entrainment when the entrained liquid is far less viscous than the
exterior, which show that interface shape at $Q_c$ can be made
singular via changes in the boundary conditions~(\cite{Zhang04}). 

Finally, the robustness of the logarithmic coupling even under drastic 
changes in the withdrawal conditions suggests that (\ref{exp_form}) in fact
corresponds to a generic feature of selective withdrawal from two
layers of equal viscosity. This leads us to re-examine experimental
measurements of $\kappa$ and $h$, previously interpretted in terms of
a continous evolution towards a change in interface topology. In the next
subsection we compare experimental results against numerical results
obtained using the simplest model of withdrawal and show that the
logarithmic coupling (\ref{exp_form}) is also a good description of
the steady-state interface evolution in the experiments.

%% file: compare.tex

Here we compare three key results from the numerics against the
measurement previously obtained by Cohen \& Nagel~(\cite{CohenNagel02}). 
First, we analyze the $\tilde{h}(\tilde{Q})$ obtained in the experiment and 
show that the measurements are consistent with the interpretation that
$h_c-h$ scales with $\sqrt{Q_c - Q}$ near the transition. 
Measurements of the hump
curvature also agree well with the calculated values and begin to saturate as
$Q_c$ is approached. Finally we compare the measured $\kappa$ versus $h$
curves against the calculation and show they also agree. 

\begin{figure}
  \centering
  \includegraphics[width=0.75\textwidth]{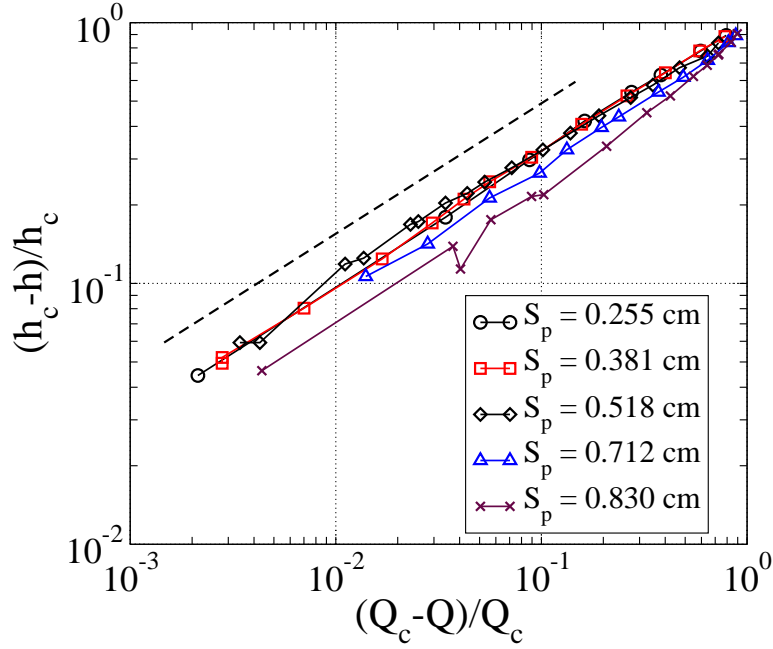}
  \caption{Rescaled hump heights $(h_c-h)/h_c$ versus
rescaled withdrawal flux $(Q_c-Q)/Q_c$ for measurements obtained with
$5$ different tube heights. The tube heights are $\tilde{S}_p= 0.260$
cm, $0.381$ cm, $0.518$ cm, $0.712$ cm and $0.830$ cm. The capillary
lengthscale in the experiment is $0.3$ cm and the layer viscosity
ratio (lower/upper) is $0.86$. The calculation results are for
$\tilde{S}/a = 0.2$ and reservoir pressure $p=0.01$.}
  \label{fig:tl_v_q_exp_sim}
\noindent\rule{5.3in}{.5mm}
\end{figure}

To make the comparison, we chose measurements from five different
experiments, spanning the full range of tube heights used. 
Figure~\ref{fig:tl_v_q_exp_sim} shows how the measured hump height
saturates as $Q$ approaches $Q_c$. As was done 
with the numerical results, in generating
figure~\ref{fig:tl_v_q_exp_sim} we allowed ourselves to vary
$h_c$ and $Q_c$ values within the experimental error bars,
which are about $5 \%$, in order to generate the best power-law fits
for the measurements. For four sets of the data, the saturation
behavior is completely consistent with a square-root scaling with
$\delta q$. The
set with the largest tube height ($S_p = 0.830$ cm) shows a slight
difference in the scaling behavior. We do not fully understand
the origin of this discrepancy. We speculate that
withdrawal experiments at
larger tube heights require running the pump at higher flow
rates. This may create more noise and/or signficant inertial effects,
hence resulting in a discrepancy with experiments performed at lower
tube heights. Overall the agreement shows that the hump height in the
experiment experiences a saddle-node bifurcation at $Q_c$. 

\begin{figure}
  \centering
  \includegraphics[width=0.7\textwidth]{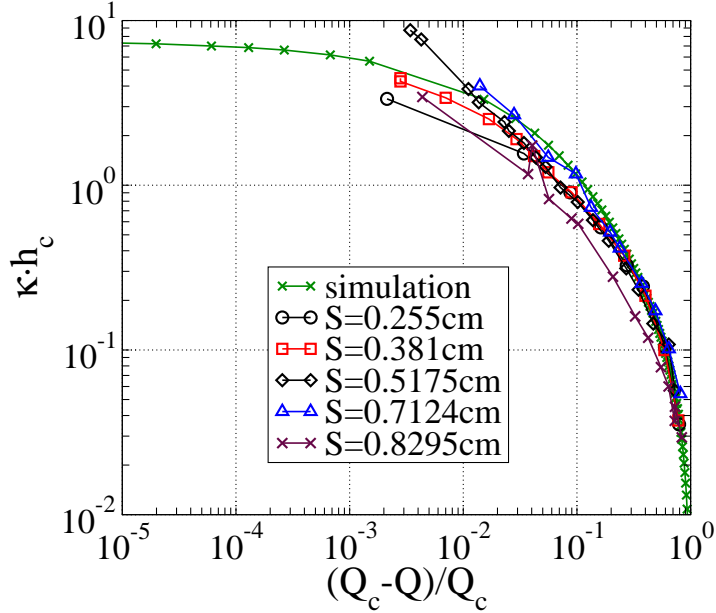}
  \caption{Tip curvature versus rescaled flow rate $\left(Q_c-Q\right)/Q_c$ for
the numerical model and experimental data. The tip curvature has been
non-dimensionalized by $h_c$. Range of the numerical results have been
truncated to display the comparison clearly.}
  \label{fig:tc_v_dqqc_exp_sim}
\noindent\rule{5.3in}{.5mm}
\end{figure}

Next we compare measurements of the hump curvature against calculated
values. Since the hump curvature satures at a much smaller 
value of $\delta q$, below the dynamic range of the experiment, we
cannot compare the saturation dynamics, which is independent of the 
details of the numerical model, directly. Instead we compare
the $\tilde{\kappa}(\delta q)$ curves obtained experimentally against
the results obtained using our minimal numerical model, described in
section 3.1, with $S \equiv \tilde{S}/a = 0.2$. This is done in
figure~\ref{fig:tc_v_dqqc_exp_sim}. The $Q_c$ values which produced
the best power-law fit for the experimental data 
in figure~\ref{fig:tl_v_q_exp_sim} are used in 
figure~\ref{fig:tc_v_dqqc_exp_sim}. (More information on the
difference between our analyses and those performed in the original
papers~(\cite{CohenNagel02}) can be found in the 
appendix) As was done for numerical results
obtained with different sink heights, we account for the change in the
absolute size of the hump at different tube heights by rescaling
$\tilde{\kappa}$ by $h_c$, the hump height at transition. 
This causes the different curves associated with different tube
heights to collapse onto roughly a single curve. Significantly the
collapsed curve shows some evidence of saturation as $\delta q$
approaches $0$. In order to 
display the comparison clearly, the range of value for both 
$\kappa \cdot h_c$ and $(Q_c-Q)/Q_c$ have been truncated. 
The calculated curve goes through the experimental values and shows
exactly the same trend. Note that our choice of $S=0.2$ for the
numerical results is primarily a
matter of simplicity, since that is the set of results discussed in
section 4.1. From section 4.2 we have seen that 
$\kappa_c \cdot h_c$ varies only weakly with $S$, so we could have
used any $O(1)$ value for the dimensionaless sink height $S$ and
gotten good agreement in this rescaled curvature plot. 
We conclude from
the good agreement that the hump curvature also saturates in the experiment. 

\begin{figure}
  \centering
  \includegraphics[width=0.7\textwidth]{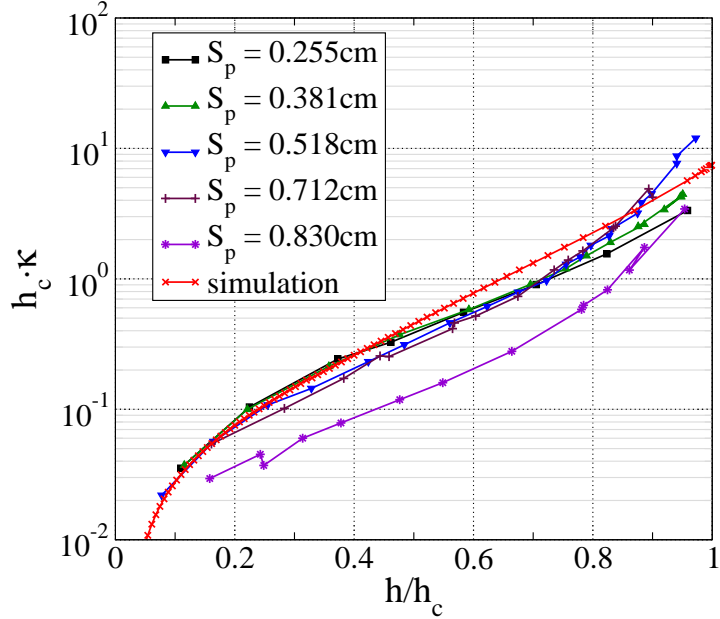}
  \caption{Comparison between measurements at five different pipette
heights and calculation. Note the range of $\kappa \cdot h_c$ has
again been truncated to display the overlap with measurements more clearly.}
  \label{fig:dh_kappa_exp_sim}
\noindent\rule{5.3in}{.5mm}
\end{figure}

Finally we plot the rescaled measured hump curvature as a function of the
rescaled hump height $h/h_c$ (Figure~\ref{fig:dh_kappa_exp_sim}). 
The measurements at the largest tube height ($S_p = 0.830$ cm) are
displaced from the four other sets. Also, the measured curves for the three,
larger tube heights show a slight but systematic upturn at large
$h/h_c$, when compared against the calculated results. 
The two data sets for the lower tube heights appear to follow a
logarithmic relation. While it is unclear what causes these
discrepancies, overall there is good agreement between the
calculated curve and the measured curves. The agreement shows that the
simple model used here captures the main features in the evolution of
the steady-state interface shape. 

Previously, the large increase in the hump curvature near
transition was interpretted in terms of the steady-state 
interface approaching a steady-state singularity as the withdrawal flux is
increased~(\cite{CohenNagel02,Cohen04}). The good agreement between the
numerical solutions and the experimental measurements show that
this large increase in hump curvature can also be understood to result from 
a weak coupling between the overall hump shape,
characterized by the hump height, and the shape of the hump tip,
characterized by the hump curvature. The simplest way to account for
all the results, numerical and experimental, obtained for
equal-viscosity withdrawal is to assume that no singular hump solution
exists for equal-viscosity withdrawal and that the logarithmic
coupling (\ref{exp_form}) is a generic feature of selective
withdrawal.  The absence of a singular solution for equal-viscosity
withdrawal would explain why even
drastic changes in the withdrawal conditions failed to introduce
qualitative changes in the $\kappa$ versus $h$ curves calculated, and why the
product $\kappa_c \cdot h_c$ characterizing the sharpness of the hump
profile always remains below $20$. The saddle node structure of the transition 
and the logarithmic coupling between $h$ and $\kappa$ also account in
a natural way for the two features which appear puzzling in the context of
a nearly continuous shape transition: the apparent 
transition cut-off in the hump evolution and the large changes in the hump 
curvature relative to changes in hump height.  

While our results simplify the view of the steady-state interface
evolution significantly, they do not offer a full explanation. We do
not know why the logarithmic coupling should
arise in selective withdrawal in the first place. Section 5 offers 
a brief, and speculative,
discussion of its possible origin, focusing in particular on
a comparison between previously proposed singular solutions and the
logarithmic relation found in our numerical solution.

%% file: discuss.tex

A natural way to reconcile the 
logarithmic evolution of the hump height relative to hump 
curvature~(\ref{exp_form}) obtained here 
with the previously proposed
idea that the sharpening of the steady-state hump tip is associated
with an approach towards a steady-state singularity is to suppose that
the interface evolution is indeed organized by the existence of a
steady-state singularity, as suggested by 
previous studies. However, the singular shape is not nearby in
the sense that it is attained at $Q$ just above $Q_c$, where a
transition from selective withdrawal to viscous entrainment takes
place. Instead, the singular solution corresponds to an isolated
singularity, existing at a value $Q_*$ much
larger than $Q_c$. Correspondingly, the height of the singular hump
$h_*$, is much larger than $h_c$, the hump height at transition.
In other words, the steady-state interface evolves towards a singular
shape at $Q_*$ as the withdrawal flux is increased. However, this
evolution is cut-off at a withdrawal flux $Q_c$ far below $Q_*$ by a
saddle-node bifurcation.
While unusual, a steady-state cusp solution corresponding to an isolated
singularity has been found for 2D selective withdrawal of inviscid
fluid~(\cite{tuck84}) so this is a possible scenario for the
steady-state evolution.  

Under these assumptions, the power-law scaling 
\beq
\frac{\kappa}{C} = \left( \frac{h_*}{h_*-h} \right)^\beta
\label{beta}
\eeq
where $C$ is a characteristic curvature scale, indeed 
assumes an approximate
logarithmic form. Since $h_c \ll h_*$, 
the ratio $h/h_*$ is small over the entire range. 
Taylor series expansion of the natural log of (\ref{beta}) yields 
\beq
\ln \left( \frac{\kappa}{C} \right ) = 
-\beta \ln \left( 1 - \frac{h}{h_*} \right) = 
\beta \left( \frac{h}{h_*} + \frac{1}{2} 
\left( \frac{h}{h_*} \right )^2 + \cdot \cdot \cdot \right ) \ . 
\label{logrelation} 
\eeq
When $h_c/h_*$ is sufficiently small, the higher order 
terms are negligible, and we obtain a logarithmic relation between
$\kappa$ and $h$
\beq
\ln \left(\frac{\kappa}{C} \right) = \left( \frac{\beta
h_c}{h_*} \right) \frac{h}{h_c} \ . 
\label{betalog}
\eeq

According to Cohen \& Nagel, the singular
shape is characterized by a scaling exponent $\beta$ between $0.75$
and $0.82$~(\cite{CohenNagel02}). In the long-wavelength analysis of
spout 
formation~(\cite{Zhang04}), Zhang found that the steady-state 
spout shape approaches a singular, conical shape. This would
correspond to $\beta = 1$. More recently, experimental and scaling
analysis 
by Courrech du Pont and Eggers show that the cusp observed
in a viscous drainage experiment approaches a conical
shape~(\cite{sylvain06}), with $\beta=1$. Given these values,
consistency requires that 
the coefficient $\beta h_c/h_*$ must be much smaller than $1$, 
since (\ref{betalog}) is derived 
under the assumption that $h_c/h_* \ll 1$.  We can check this against
the numerical results by fitting the calculated $\kappa$ versus $h$
curves with the form 
\beq
\ln \left ( \frac{\kappa}{\kappa_0} \right ) = b \left ( \frac{h}{h_c}
\right ) 
\eeq
where $\kappa_0$ is the extrapolated intercept at $h=0$ and the slope
$b = \beta h_c/h_*$ in (\ref{betalog}). Fits to the calculated
curves for $S=0.05$, $0.2$ and $50$ yield, respectively, 
$b$ values $5.5$, $5.9$, and
$1.9$, all above $1$. Consequently, $h_c/h_*$ is $O(1)$ which
invalidates  
the Taylor expansion in equation~(\ref{logrelation}). This clearly rules
out  
the possibility that the logarithmic relation reflects the existence 
of an isolated singular hump solution for equal-viscosity withdrawal. 

A number of future studies would complement this one. It
would be useful to explicitly calculate the unstable hump
solution, given its relation with the stability of the hump
solutions. In light of recent
works~(\cite{chaieb_preprint,Zhang04,sylvain06})
suggesting that the steady-state interface indeed evolves towards a
singular shape when the lower layer is much less viscous than the
upper layer, it would also be worthwhile to analyze the selective
withdrawal regime when the two liquid layers have unqual
viscosities. We expect that the final transition from selective
withdrawal to viscous entrainment occurs via a saddle-node
bifurcation even when the liquid layers have unequal
viscosities. However, the limiting situation where the lower layer
has a vanishingly small viscosity relative to the upper layer
corresponds to a qualitative change in the entrainment dynamics 
since only the flow in the upper layer is significant. This then
suggests a qualitative change in the steady-state shape evolution as well.

%% file: conclude.tex

In conclusion, this paper presents a numerical study of a simple
example of flow-driven topological transition: 
the transition from selective withdrawal to viscous entrainment. To
uncover the fundamental nature of the topology change, 
we focus on the selective withdrawal regime, and analyze how 
the steady-state shape of an interface between
two immiscible and viscous liquid layers changes as the imposed
flow strength increases.  We analyze a model problem in which the 
interface deformation is solely controlled by viscous stresses and
surface tension. Its results show that the hump solution
corresponding to the selective withdrawal regime experiences a
saddle-node 
bifurcation at a threshold flow rate. Above the threshold, no hump
solution exists. As observed in the experiments, the hump curvature
increases dramatically near transition while the hump height increases
only weakly, but the increase corresponds to a logarithmic coupling
between the hump height and the hump curvature. This suggests that the
evolution of the hump tip remains weakly coupled with the evolution of
the overall hump shape as the transition is approached. Moreover, we
find that this logarithmic coupling is robust. Numerical
solutions for different realizations of viscous withdrawal, as well as
measurements, all show a logarithmic coupling between hump height and
hump curvature.

%% file: appendix.tex
As was seen in previous sections, compared to the hump height, the tip curvature
shows evidence of saturation towards a final value only when $Q$ is very close
to $Q_c$. As a consequence, when the entrainment threshold is approached, the
tip curvature can appear to diverge while the hump height reaches a saturation
value. In particular, when $\kappa$ is plotted against $(Q_c-Q)/Q$ instead of
$(Q_c-Q)/Q_c$, it is very difficult to distinguish between a continued power law
and a turn over into saturation with the available range of experimental data.
This is because even a slight shift in the value of $Q_c$ can straighten out the
curves. The apparent power law in the $(Q_c-Q)/Q$ range between $1$ and
$10^{-2}$ seems to be an artifact of the fact that, when $Q$ is small,
the denominator of $\left(Q_c-Q\right)/Q$ is close to $Q_c$ and almost
constant, thus the entire expression scales roughly like $1/Q$. Furthermore, for
small flow rates $Q$ the system response to the outside flow can be assumed to
be almost linear, with a linear dependence of $\kappa$ on $Q$. This would
then asymptotically give rise to a powerlaw with an exponent of $-1$.
\begin{figure}
  \centering
{
  \includegraphics[width=0.7\textwidth]{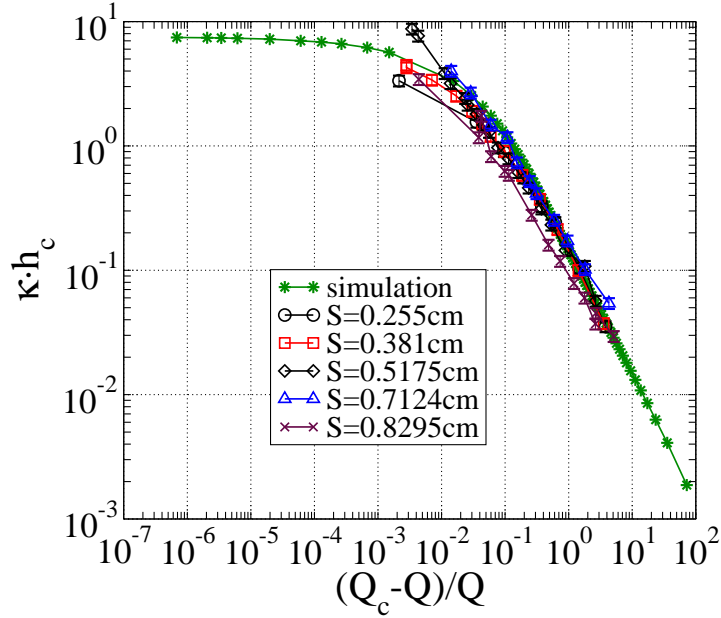}
}
  \caption{Tip curvature versus rescaled flow rate $\left(Q_c-Q\right)/Q$ for
the numerical model and experimental data. The tip curvature has been
non-dimensionalized using the threshold hump height $h_c$. Errors in the
experimental tip curvature were estimated to be approximately 10\%. The model
system
parameters are $S=0.2$, $p_0=0.01$.}
\label{fig:tc_v_dqq_exp_sim}
\noindent\rule{5.3in}{.5mm}
\end{figure}